\documentclass[aps,prx,reprint,
superscriptaddress,floatfix,longbibliography,footinbib]{revtex4-1}
\usepackage{mathtools}
\usepackage{amsfonts}
\usepackage{amsmath}
\usepackage{float}
\usepackage{sidecap}
\usepackage{bbold}
\usepackage[version=4]{mhchem}
\usepackage[doipre={DOI:~}]{uri}
\usepackage{dcolumn}
\newcommand{\nocontentsline}[3]{}

\usepackage{xcolor}

\newcommand{\be}{\begin{equation}}
\newcommand{\ee}{\end{equation}}
\newcommand{\bea}{\begin{eqnarray}}
\newcommand{\eea}{\end{eqnarray}}
\newcommand{\ben}{\begin{equation*}}
\newcommand{\een}{\end{equation*}}
\newcommand{\ba}{\begin{align}}
\newcommand{\ea}{\end{align}}

\newcommand{\mbf}{\mathbf}

\begin{document}

\title{A high-cooperativity confocal cavity QED microscope}

\author{Ronen M.~Kroeze}
\affiliation{Department of Physics, Stanford University, Stanford, CA 94305, USA}
\affiliation{E.~L.~Ginzton Laboratory, Stanford University, Stanford, CA 94305, USA}
\author{Brendan P.~Marsh}
\affiliation{E.~L.~Ginzton Laboratory, Stanford University, Stanford, CA 94305, USA}
\affiliation{Department of Applied Physics, Stanford University, Stanford CA 94305, USA}
\author{Kuan-Yu~Lin}
\affiliation{Department of Physics, Stanford University, Stanford, CA 94305, USA}
\affiliation{E.~L.~Ginzton Laboratory, Stanford University, Stanford, CA 94305, USA}
\author{Jonathan Keeling} 
\affiliation{SUPA, School of Physics and Astronomy, University of St. Andrews, St. Andrews KY16 9SS, United Kingdom}
\author{Benjamin L.~Lev}
\affiliation{Department of Physics, Stanford University, Stanford, CA 94305, USA}
\affiliation{E.~L.~Ginzton Laboratory, Stanford University, Stanford, CA 94305, USA}
\affiliation{Department of Applied Physics, Stanford University, Stanford CA 94305, USA}

\date{\today}

\begin{abstract}

Cavity QED with cooperativity far greater than unity enables high-fidelity quantum sensing and information processing. The high-cooperativity regime is often reached through the use of short, single-mode resonators. More complicated multimode resonators, such as the near-confocal optical Fabry-P\'{e}rot cavity, can provide intracavity atomic imaging in addition to high cooperativity.  This capability has recently proved important for exploring quantum many-body physics in the driven-dissipative setting. In this work, we show that a confocal cavity QED microscope can realize cooperativity in excess of 110. This cooperativity is on par with the very best single-mode cavities (which are far shorter) and 21$\times$ greater than single-mode resonators of similar length and mirror radii. The 1.7-$\mu$m imaging resolution is naturally identical to the photon-mediated interaction range.  We measure these quantities by determining the threshold of cavity superradiance when small, optically tweezed Bose-Einstein condensates are pumped at various intracavity locations. Transmission measurements of an ex situ cavity corroborate these results.  We provide a theoretical description that shows how cooperativity enhancement arises from the dispersive coupling to the atoms of many near-degenerate modes.

\end{abstract}

\maketitle

\section{Introduction}

\begin{figure}[t!]
\centering
\includegraphics[width=\columnwidth]{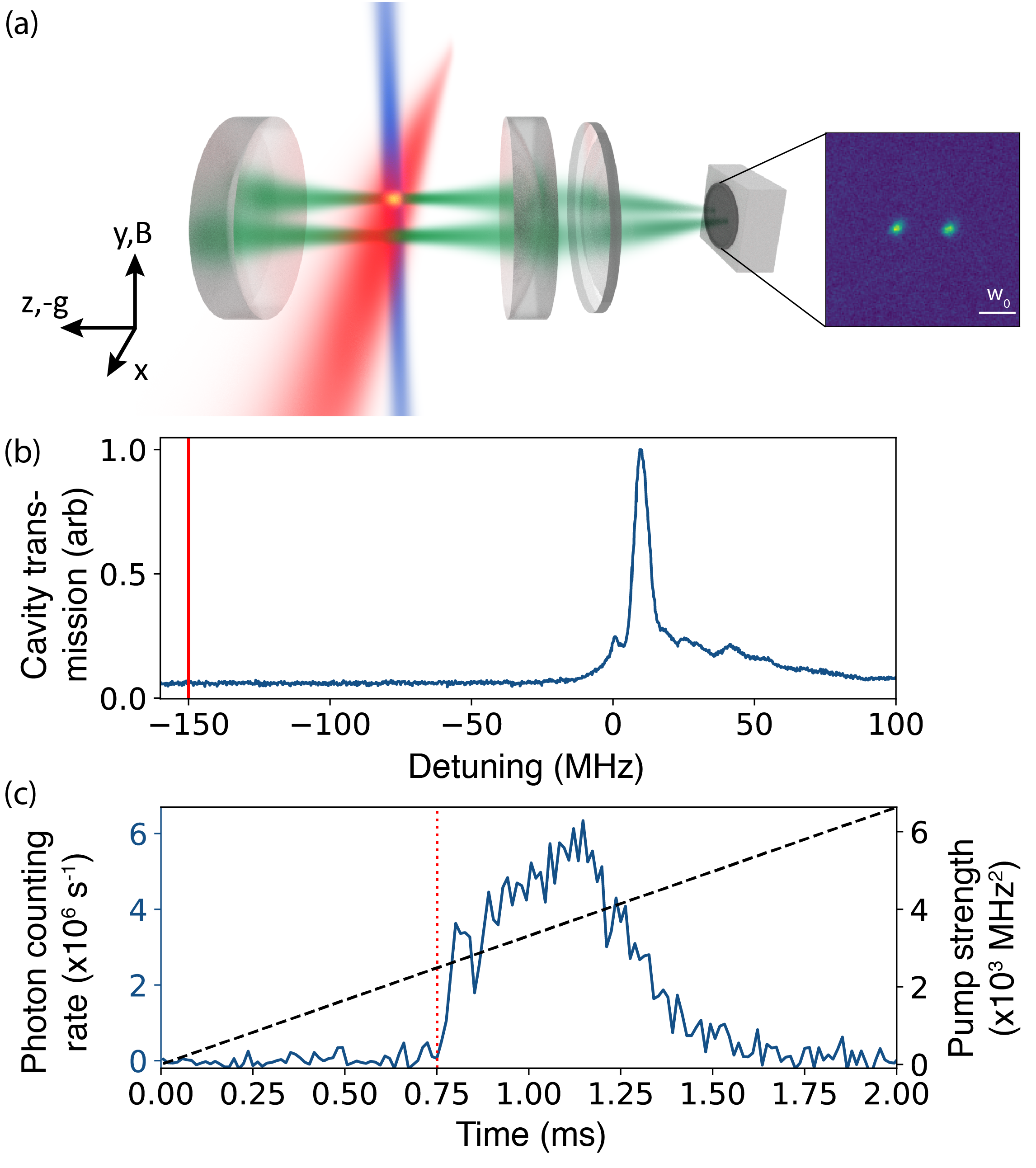}
\caption{(a) Illustration of the experiment. A $^{87}$Rb BEC (orange) is positioned using an optical tweezer dipole trap (blue) inside a confocal cavity. It is pumped by a transverse, running-wave laser (red), which induces a superradiant self-organization phase transition above a critical pump power. This scatters light into a synthetic mode---i.e., a near-degenerate mode superposition---of the cavity (green), which can be imaged onto a detector (gray box). The inset image shows an example of the synthetic mode emission; two spots correspond to the local and mirror components of the field. The waist of the fundamental mode of the cavity is $w_0=35$~$\mu$m. (b) Transmission spectrum of the employed confocal cavity taken by using a probe beam focused at the cavity center with waist of 25~$\mu$m (at the cavity midplane). The red line indicates a typical transverse pump detuning from a reference point within the mode spread $\omega_\text{B}$, here chosen to be the first local maximum in the transmission. (c) Example cavity emission (solid blue) detected by a single photon counter versus time.  The pump power is shown as a dashed black line. Threshold power is indicated by the red dotted line. Superradiant emission is observed once threshold is reached and persists for ${\lesssim}1$~ms, limited by the excitation of atomic motion by the running-wave pump.\label{fig1}}
\end{figure}

Strong light-matter interactions allow for the coherent exchange of quantum information between photonic and atomic degrees of freedom. Because exchange can occur before decoherence from dissipation, strong coupling is a key ingredient in many quantum information and sensing platforms~\cite{haroche2006exploring,Northup2014qit,Reiserer2015cqn,Borregaard2017oat,Blais2021cqe}.  Moreover, it also allows photonic excitations to be coherently exchanged among atoms in such a way as to mediate particle or spin interactions~\cite{Gopalakrishnan2009eca,Gopalakrishnan2011fag,Gonzalez-Tudela2015svl,Vaidya2018tpa,Guo2019spa,Guo2019eab,Muniz2020edp,Periwal2021pia,Li2022csa,Zhang2022ass}. These can induce nonequilibrium quantum many-body phases~\cite{Mivehvar2021cqw}. 

Optical cavity QED provides the means to generate strong coupling through the interaction of an atom's electric dipole with the electromagnetic field confined as a cavity mode~\cite{Kimble1998sio}. The larger the field, the stronger the coupling.  Single-mode resonators are typically employed to enhance the field at the atom.  However, multimode degenerate cavities are of increasing interest~\cite{Kollar2015aac,Guo2021aol}.  They accommodate many modes at (nearly) the same resonant frequency~\cite{Siegman1986l} and thereby enable light-matter coupling involving the participation of spatially distinct modes. A wealth of exotic phenomena can result, including (interacting) photonic Laughlin states~\cite{Clark2020ool}, optical mode conversion~\cite{Baum2022omc}, and compliant optical lattices~\cite{Guo2021aol} arising from emergent crystallinity~\cite{Gopalakrishnan2009eca,Gopalakrishnan2010aco,Guo2021aol}.  These arise from interactions of tunable range and connectivity~\cite{Vaidya2018tpa,Guo2019spa} as well as from broken continuous symmetries induced by the cavity coupling~\cite{Guo2019eab}.  Tunable connectivity can also be realized using multifrequency-driven single-mode cavities~\cite{Torggler2014aml,Torggler2020sac,Periwal2021pia}, and a $U(1)$ symmetry may be broken in ring~\cite{Gopalakrishnan2009eca,Gopalakrishnan2010aco,Mivehvar2018dsi,Schuster2020spo} or crossed single-mode cavities~\cite{Leonard2017sfi,Leonard2017mam}. Other applications such as the simulation of synthetic pairing~\cite{Rylands2020ppt} and dynamical gauge fields~\cite{Ballantine2017mef}, spin glasses~\cite{Gopalakrishnan2011fag,Strack2011dqs,Buchhold2013dqs,Erba2021sgp},  and neuromorphic optimizers (enabling, e.g., associative memory)~\cite{Gopalakrishnan2012emo,Rotondo2018oqg,Fiorelli2020soa,Marsh2021eam} may arise from the unique nonlocal interactions provided by multimode cavity QED~\cite{Vaidya2018tpa,Guo2019spa}.  

The relative strength of light-matter interactions in a single-mode cavity is captured by cooperativity $C$~\cite{Kimble1998sio,Tanji-Suzuki2011iba}. This figure-of-merit compares the single-atom, single-photon interaction strength $g_0$ to two dissipative rates through the ratio $C\equiv g_0^2/\kappa\gamma_\perp$~\footnote{$g_0$ is measured with respect to the peak field of the Gaussian TEM$_{00}$ mode at the cavity center.}. Field dissipation at the rate $\kappa$ occurs due to emission through the cavity mirrors (and absorption therein).  Atomic decoherence proceeds via free-space spontaneous emission at the rate $\gamma_\perp$. For a two-level atom, $\gamma_\perp=\Gamma/2$, where $\Gamma$ is the population decay rate. The strong coupling regime arises when $C>1$, and high-fidelity quantum state preparation usually requires $C\gg1$~\cite{haroche2006exploring}. In this regime, coherent mixing of atomic and photonic excitations is sufficiently strong to generate quantum entanglement~\cite{Raimond2001mqe}, effects such as photon blockade~\cite{Birnbaum2005pbi}, and coupling to Rydberg atoms~\cite{Jia2018asi,Chen2022hfb}. 

The cooperativity of a Fabry-P\'{e}rot cavity is inversely proportional to the transverse mode area at the atom~\cite{Tanji-Suzuki2011iba}.  A common strategy to raise $C$ involves reducing the mode waist $w_0$. This can be accomplished by shrinking the length of optical single-mode cavities to the hundred-micron regime or less. By contrast, multimode cavities can provide small effective mode waists while maintaining large cavity lengths for easy transverse access and small $\kappa$ (if desirable). This leads to an enhanced light-matter interaction strength $g$. We define the multimode cooperativity as $C_\text{mm} = g^2/\kappa\gamma_\perp$. $C_\text{mm}$ plays the same figure-of-merit role as $C$ does for single-mode cavities. 

Single-mode cavities support TEM$_{lm}$ modes that are spaced in frequency much further than any other scale.   Relevant scales include $g_0$, $\kappa$, and the detuning $\Delta_C=\omega_P-\omega_C$ between the pump $\omega_P$ and cavity $\omega_C$ fields.  By contrast, multimode cavities exhibit a large (near) degeneracy of modes within a bandwidth $\omega_\text{B} \leq \Delta_C$. Concentric cavities are a particularly simple example.  However, practicable cavities cannot reach the concentric limit at which the cavity length $L$ equals twice its mirrors' radius of curvature $R$ without becoming unstable~\cite{Siegman1986l}. By contrast, the confocal cavity ($L=R$) is stable in practice. This resonator geometry provides a narrow mode waist, even for $L$'s as large as 1~cm, as we demonstrate in this work.

We experimentally show how a confocal cavity can enable the creation of an electric field of narrow waist at the position of intracavity atoms coupled to a transverse pump field; see Fig.~\ref{fig1}a.  This provides access to the high-cooperativity limit:  The cavity enables $^{87}$Rb atoms that are pumped on the $D_2$ atomic transition to couple to light with a multimode cooperativity of $C_{\text{mm}} = 112(2)$ because the mode waist can shrink to $w_\text{eff} = 1.7(2)$~$\mu$m~\footnote{While we employ the term `mode waist,' we actually mean the HWHM of the field.  This is because the synthetic modes are not simple Hermite-Gaussians and so lack the same definition of mode waist as discussed in, e.g., Ref.~\cite{Siegman1986l}.}. As explained in detail below, this is possible because the cavity can support a localized photonic wavepacket that matches the size and position of the atomic gas. That is, the confocal cavity can superimpose the electric fields of many (near) degenerate modes into a concentrated area at the atomic position. For comparison, this waist is far smaller than the Gaussian TEM$_{00}$ waist $w_0=35$~$\mu$m of a single-mode cavity with the same $R$ and nearly the same length. If such a cavity has the same finesse, then it will yield a $C$ of only 5.2.  A cooperativity of 110 is on par with the much shorter, state-of-the-art single-mode cavities that have been employed for quantum gas cQED~\cite{Chang2018cqm,Mivehvar2021cqw}. 
In this work, cooperativity is measured through its effect on the threshold of cavity superradiance induced by transversely pumping an intracavity BEC. We qualitatively explain this effect in Sec.~\ref{howitworks} before describing the measurement scheme and experimental results on cooperativity enhancement in Secs.~\ref{scheme} and~\ref{measurements}, respectively. A study of the optical resolution of an ex situ confocal cavity of the same configuration corroborates these results and is presented in Sec.~\ref{Sec:ExSituImaging}. The thresholds are connected to cooperativity enhancement by using the theoretical description presented in App.~\ref{theory}. Appendices~\ref{fitprocedure}
and~\ref{alloptical} present details of the data fitting procedure and the ex situ cavity measurement, respectively. The estimate of the number of modes coupled to the atoms is in App.~\ref{effectivemodenumber}.

\section{Cooperativity enhancement through mode superposition}\label{howitworks}

We now describe the mechanism by which cooperativity is enhanced in a confocal cavity. Multimode cavity QED, of which the confocal configuration is one instance, allows the atoms inside the cavity to dynamically shape the electric field modes to which it couples.  Above a threshold pump power, the atoms maximize the field at their position by way of a superradiant phase transition~\cite{Kirton2018itt}.  This leads to an enhancement of light-matter coupling.  To explain this effect---and the measurements we will later report---we must first consider the Green's function of an idealized confocal cavity, one in which all mode families are perfectly degenerate~\cite{Siegman1986l}.  We will then consider the case involving practicable resonators.

\subsection{Ideal multimode cavities}

Cavity modes are often represented in a Hermite-Gaussian basis consisting of nonlocal mode functions.  We denote them by their transverse field profile $\Xi_\mu(\mbf{r})$, where the $\mu=\{l_\mu,m_\mu\}$ are transverse node indices.  However, an equally valid basis for degenerate cavities consists of local mode functions that tile the cavity midplane located at $z=0$~\footnote{The full mode function may also possess a nonlocal component that will be discussed below~\cite{Vaidya2018tpa,Guo2019spa,Guo2019eab}.}. A cavity supporting an infinite number of degenerate modes will admit  superpositions of the $\Xi_\mu(\mbf{r})$ that form delta functions at each point in the midplane.  Each element of this rediagonalized mode basis may be referred to as a supermode.  An atom couples to the supermode formed at its location $(\mbf{r},z)$.  

The steady-state cavity field $\Phi(\mbf{r},z)$ that arises due to scattering of the pump into the cavity is, to lowest order,
\be\label{eq:cavfield}
\Phi(\mbf{r},z)=\Phi_0\int d\mbf{r'} dz' \mathcal{D}(\mbf{r},\mbf{r'},z,z')\rho(\mbf{r'},z'),
\ee
where the atomic distribution is $\rho(\mbf{r},z)$ and is normalized by $\int \rho(\mbf{r},z) d\mbf{r}dz=1$, $\mathcal{D}(\mathbf{r},\mathbf{r}',z,z')$ is the cavity Green's function, and $\Phi_0$ is the scaled field strength given in Appendix~\ref{theory}.  As we show in App.~\ref{Sec:Enhancement}, the maximum multimode cooperativity $C_\text{mm}$ is set by the strength of the cavity field at the location of a point source, that is
\begin{equation}\label{eq:cenhancement}
    \frac{C_\text{mm}}{C} =\max\limits_{\mbf{r},z}\frac{\Phi(\mbf{r},z)}{\Phi_0} =\max\limits_{\mbf{r},z} \mathcal{D}(\mbf{r},\mbf{r},z,z).
\end{equation} 
For a confocal cavity, this maximum is at the cavity center and $C_\text{mm}/C=\mathcal{D}(\mbf{0},\mbf{0},0,0)$. The general expression for this Green's function is
\begin{align}\label{eq:cav_interaction}
    \mathcal{D}(\mbf{r},\mbf{r'},z,z')&=\sum\limits_\mu W_\mu\Xi_\mu(\mbf{r})\Xi_\mu(\mbf{r'})\\
    &\quad\times\cos[kz-\theta_\mu(z)]\cos[kz'-\theta_\mu(z')],\nonumber
\end{align}
where the longitudinal mode phase is $\theta_\mu(z)$ and $W_\mu$ is the weight of each mode in the superposition~\cite{Guo2019spa,Guo2019eab}.  

For an ideal single-mode cavity, $W_{00}=1$ and all other $W_\mu=0$. The transverse part of the corresponding Green's function is $\mathcal{D}_{00}(\mbf{r},\mbf{r'})\propto\exp[-(r^2+r'^2)/w_0^2]$, and the field is $\Phi(\mbf{r})\propto\exp(-r^2/w_0^2)$, regardless of $\rho(\mbf{r},z)$.  Thus, no spatial features of the gas are mapped onto the spatial dependence of the single-mode cavity field. 

On the other hand, for a perfect confocal cavity with even parity~\footnote{Confocal resonators support purely even or odd parity mode families every other half free-spectral-range~\cite{Siegman1986l}. We will consider only even mode families throughout this work.}, all $W_\mu=1$ for even $n_\mu=l_\mu+m_\mu$, while $W_\mu=0$ otherwise.  This results in the ideal confocal cavity Green's function~\cite{Vaidya2018tpa,Guo2019eab}:
\bea
\mathcal{D}(\mbf{r},\mbf{r'},z,z')\propto&U^+(\mbf{r},\mbf{r'})\cos(kz)\cos(kz')\\
&+U^-(\mbf{r},\mbf{r'})\sin(kz)\sin(kz'),
\eea
where
\be\label{interactionEQ}
U^\pm(\mbf{r},\mbf{r'})=\delta\left(\frac{\mbf{r}-\mbf{r'}}{w_0}\right)+\delta\left(\frac{\mbf{r}+\mbf{r'}}{w_0}\right)\pm\frac{1}{\pi}\cos\left[2\frac{\mbf{r}\cdot\mbf{r'}}{w_0^2}\right].
\ee
The first two terms are the local self-image and mirror image, respectively.  The latter arises due to the cavity symmetry~\cite{Siegman1986l,Vaidya2018tpa}. The third is a nonlocal term arising due to Gouy phases: We will ignore it for now because it plays only a minor role in cooperativity enhancement; see App.~\ref{theory} and Refs.~\cite{Vaidya2018tpa,Guo2019spa,Guo2019eab,Marsh2021eam} for details.

\subsection{Nonideal multimode cavities}

\subsubsection{Finite number of modes}

To consider the more physical situation of a \textit{finite} number of modes, we can impose a cutoff at $n_\mu = n_c$.  This truncates the number of supported cavity modes to approximately account for the finite solid angle subtended by each mirror.  While a hard cutoff is conceptually simple, indistinguishable results are obtained using an exponential factor $\exp(-\alpha n_\mu)$. We employ the latter since it leads to analytic expressions for the Green's function.

Every supermode is a superposition of the modes $\Xi_\mu(\mbf{r})$, weighted by $W_\mu$, that have a nonzero amplitude at $\mbf{r}$.  While no longer a delta function of infinitesimal width, the supermode waist $w_\text{eff}$ remains smaller than $w_0$ in proportion to the number of modes in the superposition.  Constructive interference increases the local electric field, thereby enhancing coupling of the atom to light.  At the cavity center, the supermode coupling strength $g$ exceeds that of the TEM$_{00}$ mode by the ratio $g/g_0$, which is also the ratio by which the vacuum Rabi splitting between upper and lower polaritons increases when probed on resonance.  This leads to a cooperativity enhancement given by Eq.~\eqref{eq:cenhancement}.

The number of $\Xi_\mu$ that contribute is maximal for the supermode formed at the cavity center and diminishes as $\mbf{r}/w_0$ becomes large due to aberrations and finite mirror aperture. Nevertheless, thousands of modes have been observed to contribute in similar confocal cavities~\cite{Kollar2015aac}.  Moreover, $w_\text{eff}$ can remain smaller than $w_0$ even at distances many multiples of $w_0$ away from the cavity center~\cite{Vaidya2018tpa}.  Modes with high finesse can be observed out to at least $n_\mu = 50$.   For simplicity, we will assume that $\kappa$ is constant for all modes, but account for the attenuation of finesse through the incorporation of the aforementioned weights $W_\mu$. 

\subsubsection{Imperfect mode degeneracy}

High-finesse multimode cavities are also not perfectly degenerate.  Inevitable mirror aberrations split mode degeneracy beyond the individual mode linewidth. This complicates the story, because the atom no longer couples to a cavity eigenmode in the form of a supermode:  The collection of modes the atom couples to are not at the same frequency and therefore do not form an eigenmode.  While the use of the term `supermode' remains common in some such contexts (e.g., in near-detuned multimode lasing and superradiance~\cite{Kapon1984sao,Kollar2017scw}), and we have employed it in the past~\cite{Vaidya2018tpa,Guo2019spa,Marsh2021eam,Guo2021aol}, we will shift to using the term `synthetic mode.' This mode-like entity is defined as the photonic component of the polariton created when the atom coherently couples to all the eigenmodes with field at its position.  The field of the synthetic mode is directly accessible through the emission of cavity light. But because it is not an eigenmode, the synthetic mode dephases after a timescale inversely proportional to the frequency spread $\omega_\text{B}$ when not being pumped.

To analyze this experimentally relevant situation, we consider an atom coupled to near-degenerate modes in the far-detuned configuration $\Delta_C \gg \omega_\text{B}$. The mode bandwidth is $\omega_\text{B}/2\pi=5$-25~MHz for the 1-cm cavities studied below.  While this is roughly two orders-of-magnitude larger than $\kappa/2\pi= 137$~kHz, enough modes merge within this bandwidth to produce a continuous spectrum; see Fig.~\ref{fig1}b.  In an imperfect confocal cavity, the spread of mode frequencies $\omega_\mu$ within $\omega_\text{B}$ is more or less random. To avoid the effects of that randomness, one can pump the system far from the near-confocal point and model the modes with a linear dispersion $\omega_\mu=\omega_{00}+\epsilon n_\mu$~\cite{Vaidya2018tpa,Guo2019eab}.

To stimulate a synthetic mode, an electric field satisfying the far-detuning condition may be introduced into the cavity by either longitudinally pumping through a cavity mirror or by transverse pumping via Rayleigh scattering off the atom.  Either way, the coherently driven intracavity field manifests as a superposition of near-degenerate modes dispersively coupled to the atom.  The cavity emits this synthetic mode at the pump frequency~\footnote{Light shifts modify this frequency, but are insignificant compared to $\kappa$ for the far-detuned system considered below.}. The notion of a single $C_\text{mm}$ value is valid only in this dispersive coupling limit because there is no unique near-resonance vacuum Rabi splitting. Rather, there is a complicated spectrum of mode mixing within the forest of nearly degenerate mode resonances.

We now arrive at the synthetic mode Green's function, which accounts for both the mode truncation and the mode dispersion within $\omega_\text{B}$.  The cutoff and linear dispersion is incorporated into the weights $W_\mu=\frac{\Delta_C}{\Delta_\mu+i\kappa}e^{-\alpha n_\mu}$; the pump detuning from the near-degenerate modes is $\Delta_\mu$. The resulting Green's function still exhibits a local self-image and mirror image term, both  with finite width. The maximum of the synthetic mode Green's function sets the cavity's cooperativity, per Eq.~\eqref{eq:cenhancement}. The finite width is set by the synthetic mode's extent, which we now discuss.

\subsection{Spatial imaging and interaction range}

The field of the synthetic mode emitted from the cavity may be imaged to provide a spatial map of the photon cloud bound to each atom~\cite{Vaidya2018tpa,Guo2019spa}. Cavity emission provides real-time access to the photonic component of the intracavity polariton. Polaritons with overlapping synthetic modes enable the local exchange of photons between nearby atoms.  This can mediate interactions among atoms, and the waist of the synthetic mode sets this interaction range~\cite{Vaidya2018tpa}.  This quantity is calculated using the cavity Green's function $\mathcal{D}(\mbf{r},\mbf{r'})$, as described in App.~\ref{theory}.

Moreover, if we view the multimode cavity as a lens system, then the minimum spatial width of the synthetic mode is its resolving capability. That is, the point spread function of this `microscope' is proportional to the cavity Green's function.  The local synthetic mode waist directly determines the resolution. We use the half-width at half-maximum (HWHM) of the field for the resolution metric because its shape does not conform to a standard functional form; see App.~\ref{theory}. Note that the resolution is always less than the mirrors' numerical aperture (NA), because the confocal cavity admits only a quarter of the free-space modes at any given resonance~\footnote{Only transversely even or odd modes and only longitudinally sine or cosine modes are supported at each degenerate resonance~\cite{Siegman1986l,Guo2019eab,Guo2019spa}.}.

We now see that the coupling enhancement due to multimode degeneracy increases cooperativity and improves imaging resolution as well as introduces an interaction length scale.  The narrowness of the synthetic mode waist may be interpreted in two ways: The scale is both an interatom interaction range and the resolving capability of the mirror system. Moreover, these lengths are naturally matched and so there is no need for higher resolution imaging.  We may consider a multimode cavity to be an unusual ``active" quantum gas microscope, in that its fields mediate local interactions among atoms while their emission provides spatial information about atomic positions. We now turn to the experiments that measure the cooperativity enhancement as well as the interaction/resolution length scale. 

\section{$C_\text{mm}$ measurement scheme}\label{scheme}

Straightforward methods to determine $C_\text{mm}$ could involve either measuring the synthetic mode light shift or spatially imaging the light emitted from the cavity. Unfortunately, neither method is practicable for our current, in situ system.  The dispersive light shift is too small and direct imaging of the emission suffers from aberrations that obscure the minimum synthetic mode spot size~\footnote{One could determine the minimum synthetic mode waist by longitudinally driving the cavity with ever smaller beam waists.  However, in our present experimental system, imaging the cavity emission does not reveal the limit of cavity performance due to aberrations from the cavity substrate and low, downstream imaging NA.}. We revisit the direct imaging method using an ex situ setup designed to circumvent some of these limitations; see Sec.~\ref{Sec:ExSituImaging}.

To measure $C_\text{mm}$ in situ, we use transverse pumping to extract $C_\text{mm}$ by observing the threshold of superradiant cavity emission. This method was first employed in Ref.~\cite{Black2003ooc} where the cooperativity enhancement was interpreted in terms of geometrical optics.  Here, we build upon an alternative description from Ref.~\cite{Vaidya2018tpa} as well as perform measurements with better spatial resolution than in that work.

To explain this method, we recall that superradiance can emerge via the phase-locking of many atomic dipoles when they are transversely pumped above a threshold power. The self-organization threshold is described by the nonequilibrium Hepp-Lieb-Dicke transition (see App.~\ref{theory} for Hamiltonian). Pumping above a critical field strength $\Omega_\text{c}$ causes the atoms to spontaneously arrange themselves (in position and/or spin~\cite{Baumann2010dqp,Kroeze2018sso,Kroeze2019dsc,Mivehvar2021cqw}) into one of two possible checkerboard patterns. This forms a phase gradient that Bragg-scatters pump photons into the synthetic mode at a rate proportional to the square of the number of intracavity atoms $N$.  Above this threshold, the system condenses into a synthetic-mode polariton whose matter component is of spin and/or density-wave character.  Below threshold, the system incoherently fluctuates between polariton states.

Threshold power scales as $\Omega_\text{c}^2 \propto \Delta_C/(NC_\text{mm})$, providing a method for measuring $C_\text{mm}$.  We will now employ a more sophisticated theoretical treatment than used in Ref.~\cite{Vaidya2018tpa} to extract $C_\text{mm}$ from $\Omega_\text{c}$ measurements via the cavity Green's function.  Parameters in the function are determined by analyzing the dependence of threshold on different BEC sizes and transverse positions within the cavity. To improve accuracy, we also include higher-order effects and contact interactions. 

We now describe how the self-organization threshold for an intracavity BEC depends on $\Phi$, and thus, how $C_\text{mm}$ can be determined through Eq.~\eqref{eq:cenhancement}.  As derived in App.~\ref{theory}, the critical Rabi strength $\Omega_\text{c}$ of the transverse pump is given by the condition
\bea\label{threshold_condition} 
E_\text{dw} = &&-\frac{Ng_0^2\Omega_\text{c}^2}{\Delta_A^2\Delta_C}\\ &&\times\Re\left\{\int d\mbf{r}\rho(\mbf{r})\frac{\Phi(\mbf{r})}{\Phi_0}+\frac{Ng_0^2}{2\Delta_A\Delta_C}\int d\mbf{r}\rho(\mbf{r})\frac{\Phi(\mbf{r})^2}{\Phi_0^2}\right\},\nonumber
\eea
where $E_\text{dw}$ is the excitation energy of the density wave (set by the atomic recoil and chemical potential) and $\Delta_A$ is the atomic detuning of the transverse pump.  This field is generated by the scattering of light off the condensate and into the cavity, given by Eq.~\eqref{eq:cavfield}. The pump frequency is always far detuned to the red of the atomic transition frequency $\omega_A$, such that $\Delta_A=\omega_P - \omega_A$ is the largest scale in the Hamiltonian. This renders spontaneous emission negligible on the timescales considered.  We also tune the system so that $\Delta_{C}<0$. 

The first term in parentheses is the atomic density overlap with the intracavity light field and is the standard threshold expression~\cite{Kirton2018itt}. For better correspondence with experiment, we also consider the second-order term.  This describes the overlap with the \textit{intensity} of the cavity light. It represents the leading-order contribution of the dispersive shift. The dispersive shift in a single-mode cavity leads to a simple renormalization of the cavity detuning, $\Delta_C \rightarrow \Delta_C - \frac{Ng_0^2}{2\Delta_A\Phi_0^2}\int d\mbf{r}\rho(\mbf{r})\Phi_{00}(\mbf{r})^2$.  However, it results in a position-dependent correction in a multimode cavity. Furthermore, the dispersive shift contributes even when $\Delta_C/2\pi\approx -100$~MHz because of the localization of light into a synthetic mode.

Thus, by observing the threshold of a BEC with measured shape $\rho(\mbf{r})$~\footnote{This is determined by trap frequency measurements and corroborated by atomic absorption imaging.}, we can fit the parameters $\epsilon$ and $\alpha$ that appear in the Green's function.  This then enables the extraction of $\mathcal{D}$, or equivalently, $\Phi$, via Eq.~\eqref{eq:cenhancement}.  Note that we have departed from the case of a single intracavity atom when considering superradiant threshold measurements.  Since cooperativity is a figure-of-merit based on the point-like coupling of light and matter~\cite{Kimble1998sio}, we must assume that the atoms are nearly collocated around $\mbf{r}$ and ignore (for now) contact interactions among them. (These interactions are incorporated in both the full theory of App.~\ref{theory} and in the data analysis.) By `nearly,' we mean the experimentally relevant situation of a compact ensemble of $N$ atoms that are maximally and symmetrically coupled to the synthetic mode whose waist is the smallest supported by the cavity mirrors.  

The general setup of the cavity as an active quantum gas microscope is illustrated in Fig.~\ref{fig1}(a). A quantum gas with some spatial distribution $\rho(\mbf{r},z)$ is placed inside the confocal cavity and illuminated by a transverse pump. We first describe the apparatus and method of quantum gas preparation before discussing the threshold measurements.

\subsection{Cavity apparatus and BEC preparation}

The cavity consists of two curved mirrors with radius of curvature $R\approx1$~cm and $L\approx R$.  The single-mode cavity field emission rate is $\kappa/2\pi=137$~kHz, and together with a free spectral range of ${\sim}15$~GHz, the cavity has a maximum mode finesse of 55,000. We have realized a single-atom coupling rate of $g_0/2\pi = 1.47$~MHz for $^{87}$Rb on the D2 line~\cite{Kollar2017scw}.   This yields a single-atom, single-mode cooperativity of $C=5.2$. 

A $^{87}$Rb BEC of approximately $4\times10^5$ atoms and Thomas-Fermi radii of $[R_x,R_y,R_z]=[11.9(3),13.2(3),7.2(1)]$~$\mu$m is created inside the cavity using laser cooling and trapping procedures detailed in Refs.~\cite{Kollar2017scw,Vaidya2018tpa,Kroeze2018sso,Guo2019spa,Kroeze2019dsc,Guo2021aol}.  A more compact BEC shape is necessary to probe the minimum synthetic mode waist.   We achieve a smaller-width BEC than used in Ref.~\cite{Vaidya2018tpa} by compressing it through the shaping of the crossed optical dipole traps (ODTs) with acousto-optical deflectors (AODs). Specifically, the BEC is compressed to the desired final shape by simultaneously ramping down the dither of the AOD drive frequency while ramping up the ODT power. The BEC probe shapes we use are either small along the transverse cavity axes, $\{R_x,R_y\}<R_z$, to mimic a point-like particle~\footnote{The extent along $\hat{z}$ is irrelevant to the cooperativity, as enhancement is only due to confinement of the synthetic mode in the transverse direction.}, or narrow in one direction, $R_x<\{R_y,R_z\}$, for improving $\hat{x}$ position resolution.

The position $\mbf{r}$ of the BEC relative to the cavity center is controlled by setting the center frequency of the AOD drive.  (The BECs are trapped close to the cavity midplane $z=0$.) The final trap shape is characterized by weakly exciting sloshing modes so that we can measure the trap frequencies through momentum oscillations observed in time-of-flight (TOF) absorption imaging.   Atom number and BEC position are also measured via TOF imaging. This position calibration is further cross-checked by measuring the superradiance threshold versus $\mbf{r}$ for a single-mode cavity, where the position dependence is directly related to the known waist of the fundamental mode, $w_0$.  (We are able to tune the length of the cavity in situ to realize a single mode or confocal configuration~\cite{Kollar2015aac}.)

Once the BEC is positioned, the power of the 780-nm transverse pump beam is linearly ramped in a few ms through the superradiant transition. This beam is set to a detuning of $\Delta_A/2\pi\approx-98$~GHz. We use a running-wave transverse pump to avoid the formation of a standing-wave lattice, because the standing-wave threshold condition is more complicated than Eq.~\eqref{threshold_condition}~\footnote{This is because the band-structure of the standing-wave optical potential causes the density-wave energy $E_\text{dw}$ to be dependent on pump power.  Consequently, the threshold condition would have to be solved self-consistently.}.  This allows us to simplify the relationship between $C_\text{mm}$ and $\Omega_c$, which is important for reducing systematic error.  

The superradiant phase transition is detected by observing the cavity emission on a single-photon counter, heralded by a sudden increase in photon flux at the transition;  Fig.~\ref{fig1}c shows an example. This allows us to extract the critical pump strength $\Omega_\text{c}$.  The threshold pump power is measured at 35 different positions of the BEC along $\hat{x}$, ranging from approximately $-17$~$\mu$m to +17~$\mu$m and passing through the cavity center.  The measurement at each position is repeated $\sim$4 times. Furthermore, this procedure is repeated for various cavity detunings and with a few different BEC gas shapes to reduce systematic error. 

The peak density of the smallest BECs employed reaches $1\times10^{15}$~cm$^{-3}$. This high density results in significant atom loss during the ramp of the transverse pump. Since the threshold condition depends on atom number, we calibrate the total atom loss by measuring atom number at various points throughout the ramp, while suppressing superradiance with increased pump detuning. This allows us to infer the exact atom number at the time of threshold. The observed interaction profile is then fitted using the threshold condition Eq.~\eqref{threshold_condition} given this atom number, the measured trap frequencies (and hence the BEC shape $\rho(\mbf{r})$ using the Thomas-Fermi approximation), the cavity detuning, and the BEC position. Appendix~\ref{fitprocedure} describes this fitting procedure in more detail.

\section{Measurement results}\label{measurements}

\begin{figure}[t!]
    \centering
    \includegraphics[width=\columnwidth]{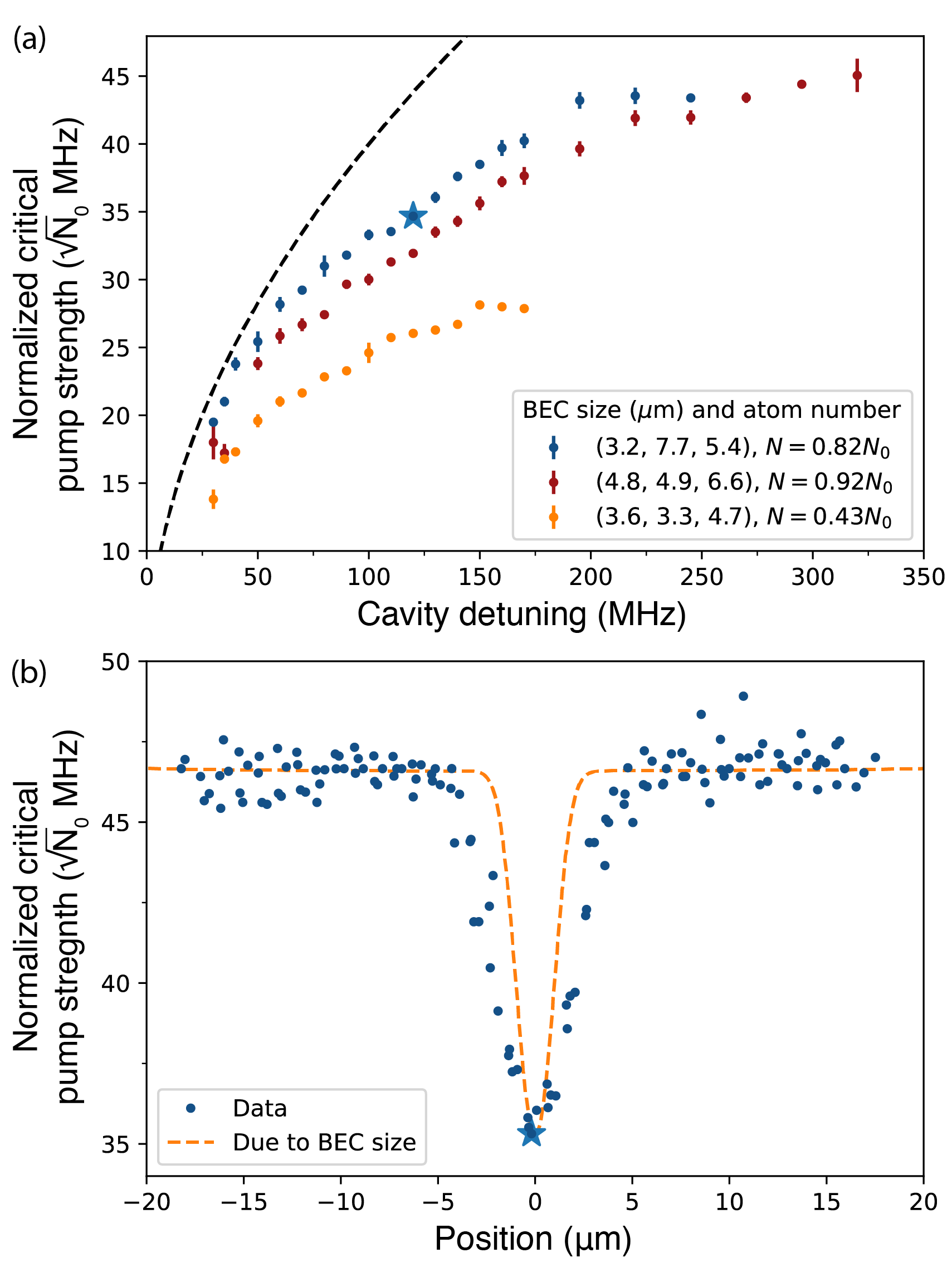}
    \caption{Threshold measurements. (a) Normalized threshold pump strength $\sqrt{N}\Omega_\text{c}$ for a BEC of three different shapes, each located at the cavity center. The pump strength is normalized by the typical BEC number, $N_0=3\times10^5$, and is plotted in units of $\sqrt{N_0}\cdot\text{MHz}$. The BEC radii are listed as $(R_x,R_y,R_z)$. The required pump power increases with pump-cavity detuning $\Delta_C$ because the photon-mediated interaction becomes weaker. However, the increase is less than in a single-mode cavity due to the participation of high-order modes. The single-mode, $\Delta_C^{1/2}$ scaling is shown as a black dashed line.  Error bars represent standard error for five shots per point. (b) Normalized threshold strength as a function of BEC position.  Data taken at $\Delta_C/2\pi=-120$~MHz with $N=0.76N_0$. The dip at the cavity center is due to the interaction enhancement from the overlap of the local and mirror fields~\cite{Vaidya2018tpa}. The minimum width of this feature is set by the size of the BEC in $\hat{x}$; the BEC probe used here has dimensions $[R_x,R_y,R_z]=[3.1,7.6,5.3]$~$\mu$m, as indicated by the orange dashed line. That the width of the dip exceeds that of the BEC indicates that the synthetic mode has a finite waist. The blue star in each panel corresponds to data taken at the same position, detuning, and trap frequencies.    \label{fig2}}
\end{figure}

We now present our measurements of threshold as a function of $\Delta_C$ and BEC probe position. 

\subsection{Threshold versus detuning and probe position}

Figure~\ref{fig2}a shows the critical pump strength as a function of $\Delta_C$ for three different shapes of a BEC held at the cavity center.  We choose to plot the critical pump strength normalized to the typical BEC atom number $N_0=3\times10^5$.  This allows us to plot all three data sets in proximity to each other despite the fact that each is derived from  experiments with different BEC population. Threshold scales with atom number as $N^{-1/2}$, so the normalized critical pump strength has units of $\sqrt{N_0}\cdot\text{MHz}$. The orange data were obtained using the smallest BEC probe we are able to make and provide the most stringent condition for placing an upper bound on the minimum synthetic mode waist supported by the cavity. 

The threshold pump strength scales as $\Omega_c\propto\Delta_C^{1/2}$ in a single-mode cavity. This scaling is illustrated in Fig.~\ref{fig2}a by a black dashed line. However, the scaling is different in a multimode cavity due to the larger number of modes through which the photons may be exchanged. As shown Fig.~\ref{fig2}a, we do indeed observe a departure from $\Delta_C^{1/2}$ scaling at larger detuning.  That is, threshold occurs at lower pump strengths in a confocal cavity.  This supports our expectation that a larger interaction energy arises due to the field contribution from these additional modes.

The maximum detuning at which we can take data is limited by both BEC heating and atom number.  The higher pump power required at large detuning leads to faster BEC heating through incoherent photon scattering. Lower initial atom numbers also require higher pump powers to reach threshold, leading to more heating.  Thus, we cannot observe threshold with the least populous BEC beyond 170~MHz (data shown in orange), while we can observe superradiance out to 320~MHz using the most populous BEC (red).

Figure~\ref{fig2}b plots threshold pump strength as a function of BEC position within the cavity.  The position of the BEC inside the cavity is scanned in a transverse plane and pumped at a fixed detuning.  This is similar to that done in Ref.~\cite{Vaidya2018tpa}, though now with a smaller-sized BEC for better resolution.  This resolution is sufficiently small that we can now discern the difference between the BEC probe width and the mode waist, as we now discuss.

The threshold strength decreases (photon-mediated interaction strength increases) at the cavity center due to the overlap of the local and mirror field components. The width of this feature is related to the minimum-detectable spot size of the local field. This measured spot size is larger than the actual minimum synthetic mode waist due to the finite size of the BEC probe.  The BEC's shape is indicated by the dashed line. We can deconvolve the BEC width from the data to obtain a measure of the minimum synthetic mode waist.  We use the HWHM of the residual minimum synthetic mode lineshape as the resolution of the cavity.  We believe that the resolution is limited by the finite extent of the remnant mode dispersion and by mirror imperfections.

\begin{figure}[t!]
    \centering
    \includegraphics[width=\columnwidth]{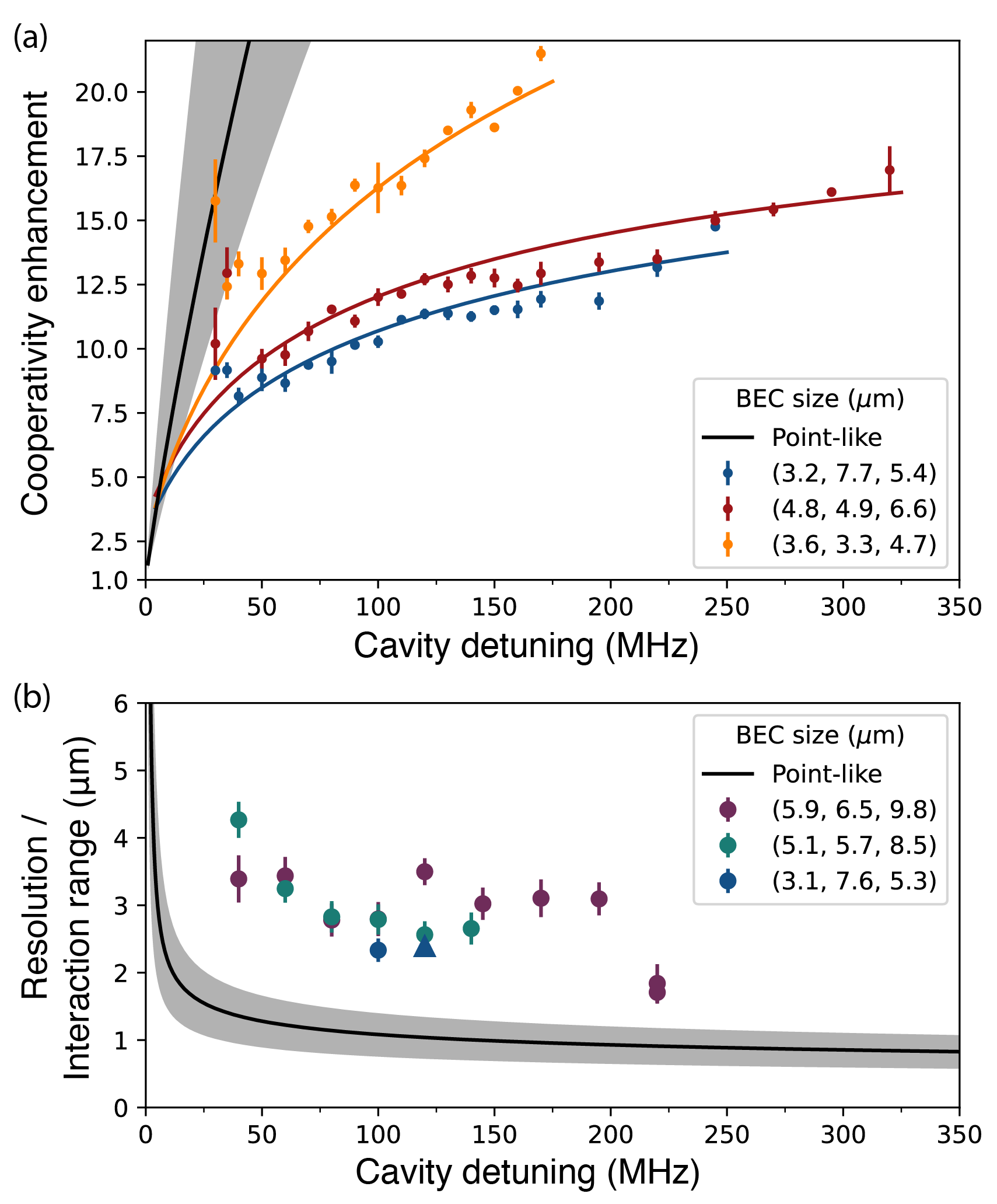}
    \caption{Cooperativity enhancement and microscope resolution/interaction range. (a) This panel presents the data in Fig.~\ref{fig2} in terms of multimode cooperativity enhancement. The enhancement increases with detuning and for narrower BECs. All solid lines are derived from the parameters $\epsilon$ and $\alpha$ obtained from a single global fit to all measurements via the threshold expression Eq.~\eqref{threshold_condition}. The black curve indicates the enhancement that a point particle would experience at the cavity center.  The gray band denotes uncertainty in $\epsilon$ and $\alpha$. The enhancement would reach 96(49) at a detuning of 300~MHz. The maximum directly measured enhancement is 21.5(3). Error bars represent standard error for five shots per point.  (b) Resolution of the multimode cavity, at the cavity center, measured by the HWHM of the synthetic mode profile; see Fig.~\ref{fig2}b for an example of a profile (convolved with the BEC probe width).  This resolution is equivalent to the photon-mediated interaction range~\cite{Vaidya2018tpa}. The data points are derived from single scans of the BEC versus position, such as that displayed in Fig.~\ref{fig2}b; see text for explanation.  The microscope reaches a minimum measured resolution of 1.7(2)~$\mu$m.  As in panel (a), the solid black line and its error band are derived from the global fit to all measurements. The blue triangle corresponds to the data presented in Fig.~\ref{fig2}b. \label{fig3}}
\end{figure}

\subsection{Cooperativity enhancement}
We next explore by how much the cooperativity is enhanced by the multimode nature of the cavity. As mentioned, this enhancement is captured by the Green's function, and in particular, its on-center value $\mathcal{D}(\mathbf{0},\mathbf{0})$. Recall that $\mathcal{D}$ is normalized to 1 in a single-mode cavity, and hence can be directly interpreted as the enhancement factor $C_\text{mm}/C$.

In Fig.~\ref{fig3}a, we show the same data as in Fig.~\ref{fig2}a, except that we scale them by $\Delta_C$ to obtain $\mathcal{D}(\mathbf{0},\mathbf{0})$. The manifest cooperativity enhancement can be qualitatively understood as the result of the modes appearing more degenerate at larger detuning. This allows the modes to form a tighter superposition and a larger electric field at the atoms' position. All solid curves result from a global theory fit for the parameters $\epsilon$ and $\alpha$.  The fit is based on datasets of eight different-shaped BECs (only three are shown here for brevity) as well as 18 scans like the one presented in Fig.~\ref{fig2}b taken at various detunings and BEC probe shapes. The complete set of data are presented in the Supplemental Material~\cite{Supp}.

The fitted theory matches well across all the datasets after applying a global detuning offset $\Delta_0$ and a scale factor to each dataset. The offset accounts for uncertainty in identifying the frequency of the lowest-order mode within the transmission spectrum, while the scale factor accounts for systematic uncertainties between datasets in the magnitude of the pump power at threshold.  All fit parameters are adjusted to minimize a global reduced $\chi^2$ metric; its minimum is 2.6. The optimal fit parameters are $\epsilon/2\pi=2.6$~MHz, $\alpha=0$, $\Delta_0/2\pi=0.8$~MHz and all scale factors are between 1.3 and 2.0.  The finite-sized BECs we use make it difficult to accurately determine the exact value of $\alpha$.  Nevertheless, we can conclude, at a 99\% confidence level, that $\alpha$ is smaller than $6\times10^{-4}$; see App.~\ref{fitprocedure} for further discussion. This implies that most of the cavity imperfection is due to the imperfect degeneracy of its modes, rather than the finite number of supported modes.

We extrapolate the BEC size down to that of a point-particle to estimate cooperativity enhancement limited solely by the cavity. A posterior likelihood distribution for the parameters $\epsilon$ and $\alpha$ is used to estimate the probability distribution for the cavity-limited cooperativity at a given cavity detuning.  From this we extract a median and 68\% confidence interval~\footnote{The distributions are not Gaussian and feature long tails. Hence, we report the median and confidence interval rather than mean and standard deviation.}. These are shown in Fig.~\ref{fig3}a as a solid black line and shaded gray region, respectively. At 100-MHz detuning, the estimated cooperativity enhancement for an ideal point particle is 42(22), and it reaches 96(49) at 300-MHz detuning.  We conservatively report $C_\text{mm}/C =21.5(3)$ from direct measurements---specifically, the orange data point at 170~MHz---when calculating the maximum multimode cooperativity $C_\text{mm}=112(2)$ that the cavity achieves. To make a rough estimate of the number of modes participating in this cavity, we use a toy model in which a finite number of modes are all perfectly degenerate. By matching the resulting cooperativity to that observed, we estimate that more than one-thousand Hermite-Gauss modes participate; see App.~\ref{effectivemodenumber} for details.

\subsection{Microscope resolution and interaction range}
We now report the minimum synthetic mode size the cavity can create. As mentioned above, this length scale may be interpreted as the photon-mediated interaction range or as the cavity microscope's resolution. 

Position scans such as that in Fig.~\ref{fig2}b provide a measurement of the cavity field profile upon deconvolution of the (known) atomic density shape.  We show how this provides the length scale of interest in App.~\ref{convolution}.  The method relies on the assumption that the profile of the cavity field has a Lorentzian shape and that the atomic density kernel is a Gaussian.  This allows us to use a Voigt profile for deconvolution. Figure~\ref{fig3}b shows the resulting resolution/interaction-range estimates at various cavity detunings and various BEC sizes.  

To estimate the HWHM of the cavity-limited resolution, we extrapolate to a point-particle by using the likelihood distribution of $\epsilon$ and $\alpha$ from the global fit, as was done for cooperativity. The median cavity resolution and 68\% confidence interval are shown in Fig.~\ref{fig3}b as a solid black line and shaded gray region, respectively. Comparing the individual data points derived from the deconvolution method with the solid curve from the global fit, we see that the deconvolution predicts a coarser resolution than that found from the global fit. This might be the result of the aforementioned approximations and/or a biasing effect in the deconvolution process.

At 300~MHz, the extrapolated cavity resolution is 0.9(3)~$\mu$m, corresponding to an effective imaging NA of 0.24(7). Improvements to mirror quality might improve this, but a beyond-paraxial treatment would be necessary to extract resolution when it nears $\lambda=780$~nm. We conservatively report the measured value of 1.7(2)~$\mu$m for the resolution of this cavity, rather than that extrapolated from the global fit. 

\section{Direct imaging of the synthetic mode}\label{Sec:ExSituImaging}

We supplement the above with an all-optical scheme for directly observing the cavity properties. While this is not feasible with the present BEC-cavity apparatus, we instead examine an out-of-vacuum cavity of the same configuration.  This allows us to observe cavity transmission with a larger geometric NA=0.42, compared to that of the in-situ cavity (NA=0.22). Its cavity mirrors are reflection-coated for 741~nm, rather than 780~nm, but they nevertheless possess  the same radii of curvature and have a comparable linewidth of $\kappa/2\pi = 198$~kHz and finesse $\mathcal{F} \approx 38,000$. This allows us to form a confocal cavity of length $L = 1$~cm, just like the in situ one.  We were able to tune the mirror alignment of this ex situ cavity into a more degenerate regime with smaller mode bandwidth $\omega_\text{B}$. Whereas the in situ cavity possesses a $\omega_\text{B}\approx25$~MHz, we were able to shrink this to ${\sim}5$~MHz in the ex situ cavity.

To probe the cavity, a tightly focused longitudinal pump field is injected through the substrate of one cavity mirror. The beam size at the cavity midplane is measured by removing the second substrate and imaging the waist onto a CCD camera. A minimum waist of $w_p \approx 1.7$~$\mu$m is observed, limited by aberrations from the input cavity substrate and various beam-shaping optics. We characterized these aberrations, allowing us to deconvolve their contributions; see App.~\ref{abberations}.

We then replace the second substrate to form the cavity and image the transmitted spot of light.  This direct imaging allows us to probe the cavity Green's function: The steady-state cavity field is exactly the convolution of the longitudinal pump with the cavity Green's function; see App.~\ref{alloptical} for details. The action of the Green's function manifests as the broadening of the waist of the cavity transmission after aberrative effects have been removed. This allows us to measure the Green's function through comparison to the single substrate observation.

Specifically, images acquired in transmission are fitted with 2D-Gaussian profiles and the Gaussian width of the transmission spot along the principal axes are extracted. The Green's function width is then estimated by the convolution relation given in Eq.~\eqref{eq:greenquadraturesum} of App.~\ref{alloptical}. The known pump waist $w_p$ is then subtracted to determine the width of the cavity Green's function, which is finally converted into a HWHM measure of the multimode cavity resolution.

\begin{figure}[t!]
    \centering
    \includegraphics[width=\columnwidth]{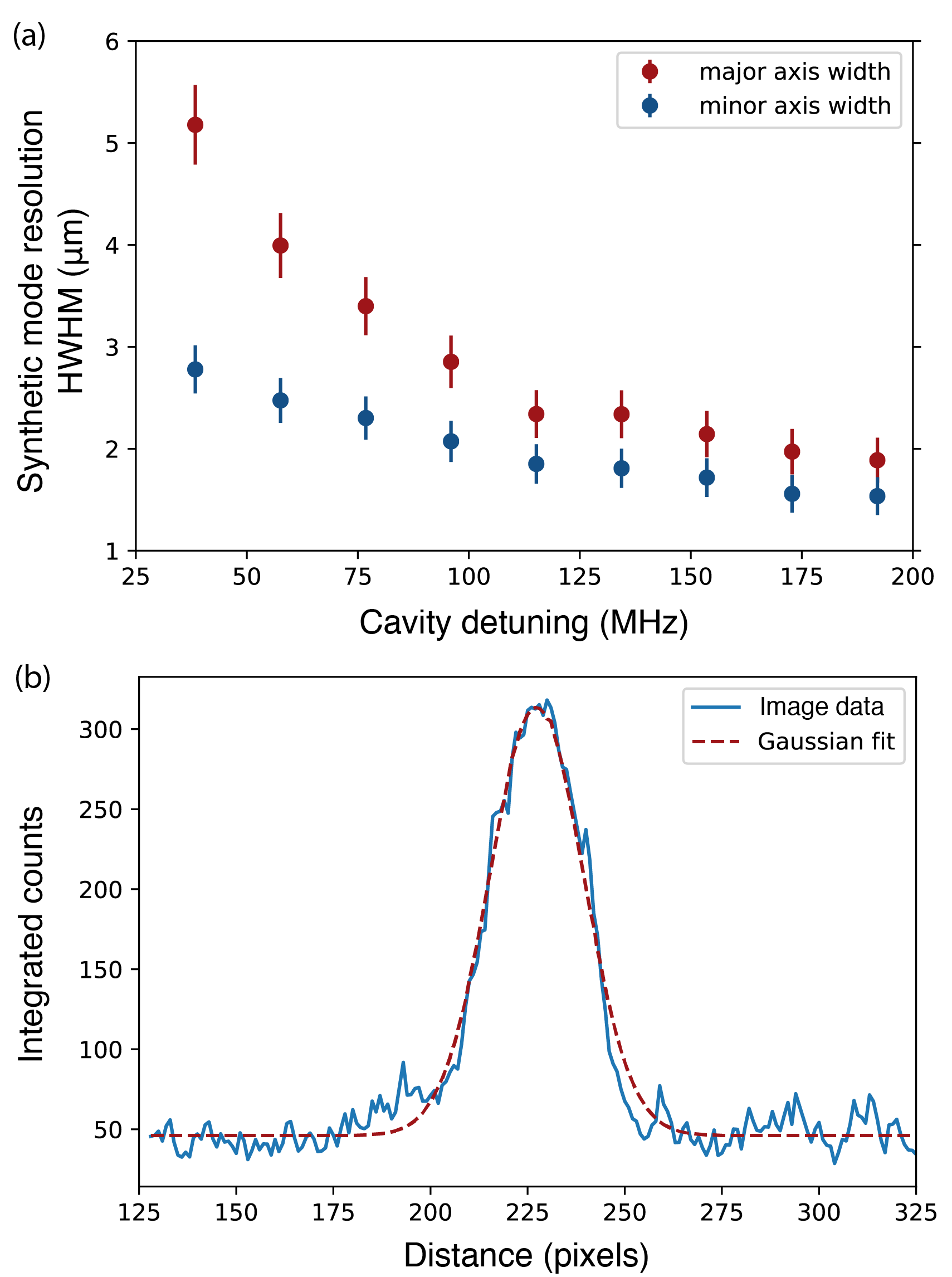}
    \caption{All-optical characterization of confocal cavity resolution. (a) Estimated widths of the cavity Green's function along the major and minor astigmatic axes versus cavity detuning. (b) Example of a typical Gaussian-like intensity profile of the cavity transmission projected onto the minor axis when the probe beam is detuned by $\sim$155~MHz.  }
    \label{fig:cavitygaussianwidth}
\end{figure}

The results versus cavity detuning are presented in Fig.~\ref{fig:cavitygaussianwidth}a. We observe a difference between the widths in the major axis (red points) and minor axis (blue points) of the 2D Gaussian. This difference is likely due to differential strain aberrations caused by the mirrors' mounts. The width of the Green's function becomes narrower as detuning increases, in qualitative agreement with the atomic superradiance threshold measurements. We note that the measured HWHMs only serve as upper bounds.  They  are limited by the aberrations and finite NA of the post-cavity imaging optics, including the output mirror substrate. The smallest observed width is ${\sim}2$~$\mu$m, which is similar to the threshold measurement results.  The resolution of this cavity may actually be lower because we were able to reduce the bandwidth $\omega_\text{B}$ by a factor of five by careful mirror alignment in this ex situ test setup.  Future work will attempt to place a tighter bound on the minimum cavity resolution and interaction achievable.

\section{Conclusion}
We demonstrated how a transversely pumped confocal cavity can enhance cooperativity through multimode superposition.  In addition to inducing strong light-matter coupling, it serves as a microscope for imaging intracavity photon-coupled atoms with a minimum waist of $1.7(2)$~$\mu$m.  This is directly matched to the cavity-induced interaction length scale since they arise from the same cavity field. Thus, as an active quantum gas microscope, this confocal cavity QED system yields a high cooperativity of 112(2), while also providing images of intracavity polaritons at their intrinsic interaction length scale.

These results support the notion that pumped confocal cavities can provide a basis for simulating strongly interacting systems such as spin glasses~\cite{Marsh2021eam}, while providing a unique, nondestructive channel for observing the microscopic behavior of these systems. An example of such usage has already been demonstrated by imaging a phonon in a dynamical optical lattice realized within this same cavity~\cite{Guo2021aol}.

\begin{acknowledgments}
We thank Sarang Gopalakrishnan and Yudan Guo for stimulating discussions. We acknowledge funding support from the Army Research Office, NTT Research, and the QNEXT DOE National Quantum Information Science Research Center.  B.M.~acknowledges funding from the Stanford Q-FARM Graduate Student Fellowship and the NSF Graduate Research Fellowship.
\end{acknowledgments}

\appendix

\section{Theoretical description of the multimode superradiant transition for a running-wave pump}\label{theory}

\subsection{Threshold condition}

We present a theoretical description of the multimode superradiant transition for the case of the running-wave pump field employed in this work.  It follows closely the theory based on standing-wave pumps presented in Ref.~\cite{Guo2019eab}, with the additional difference that we specialize to the case of a confocal resonator rather than generalize to arbitrary multimode Fabry-Perot cavities.

A system of $N$ atoms with BEC wavefunction $\Psi(\mbf{x})$, normalized as $\int d^{3}\mbf{x}|\Psi(\mbf{x})|^2=1$, placed inside the cavity is described by the Hamiltonian
\begin{widetext}\label{hamiltonian}\be\begin{split}
&H=-\sum\limits_\mu\Delta_\mu \hat{a}_\mu^\dagger\hat{a}_\mu+H_A + H_{LM}\\
&H_A = N\int d^{3}{\mbf{x}}\Psi^*(\mbf{x})\left(-\frac{\nabla^2}{2m}+V(\mbf{x})+\frac{1}{2}NU_0|\Psi(\mbf{x})|^2\right)\Psi(\mbf{x})\\
&H_{LM} = \frac{N}{\Delta_A}\int d^{3}{\mbf{x}}\Psi^*(\mbf{x})|\hat{\phi}(\mbf{x})|^2\Psi(\mbf{x}).
\end{split}\ee\end{widetext}
The modes are indexed by $\mu$ with pump-detuning $\Delta_\mu$.  The photon annihilation operators are $\hat{a}_\mu$. $V(\mbf{x})$ describes the externally applied trapping potential, which is considered to be an anisotropic harmonic trap. The atomic interaction strength $U_0=4\pi\hbar^2a/m$  is proportional to the $s$-wave scattering length $a=100a_0$ for $^{87}\text{Rb}$. The atomic excited state has been adiabatically eliminated.  $H_{LM}$ is the Stark shift due to the total light field
\begin{equation}\begin{split}\label{eq:lightfield}
\hat{\phi}&=\Omega e^{ik_rx} + \hat{\Phi}(\mbf{r}),\\
\hat{\Phi}(\mbf{r})&=g_0\sum\limits_{\mu}\hat{a}_\mu\Xi_\mu(\mbf{r})\cos\left[k_r\left(z+\frac{r^2}{2R(z)}\right)-\theta_\mu(z)\right].
\end{split}\end{equation}
Here, $\Omega$ is the Rabi strength of the transverse pump, and $k_r=2\pi/\lambda$ is the recoil momentum in terms of the light wavelength $\lambda$. $\hat{\Phi}$ is the total cavity field, and the orthogonal Hermite-Gauss transverse mode functions are
\be
\Xi_\mu(\mbf{r})=\frac{w_0}{w(z)}H_l\left(\frac{\sqrt{2}x}{w(z)}\right)H_m\left(\frac{\sqrt{2}y}{w(z)}\right)\exp\left[-\frac{r^2}{w(z)^2}\right].
\ee
The single-mode Gaussian beam size is $w(z)=w_0\sqrt{1+(\frac{\lambda z}{\pi w_0^2})^2}$, with $w_0$ the waist at the cavity midplane. The wavefront curvature is $R(z)=z+\frac{\pi^2w_0^4}{\lambda^2z}$, and $\theta_\mu(z)$ is a phase shift defined such that each of the modes satisfies the mirror boundary conditions while also accounting for the Gouy phase. The atoms are centered at longitudinal position $z_0$ and transverse position $\mbf{r}_0$. Since the atoms are positioned close to the midplane $z=0$,  $z_0\ll\frac{\pi w_0^2}{\lambda}$ and we can ignore the wavefront curvature $R(z_0)\to \infty$. The resulting $\theta_\mu(z_0)\approx n_\mu\pi/4$ gives rise to two orthogonal longitudinal quadratures available for atoms to couple. The effect of these quadratures appear only through a nonlocal interaction. To eliminate the effect of the nonlocal interaction that appears in Eq.~\eqref{interactionEQ}, we choose transverse positions $|\mbf{r}_0|< w_0\sqrt{\pi}/2$.  This allows us to focus our analysis on one of the two quadratures. See Ref.~\cite{Guo2019eab} for a more extensive discussion of the physics arising from these longitudinal quadratures.

To describe the behavior around the superradiant transition, we must describe only single scattering events.  There are two: Scattering a photon from the pump into the cavity, or scattering a photon from the cavity back into the pump. To account for these in a running-wave configuration we make the following ansatz for the atomic wavefunction:
\be\label{eq:atomic_wf}\begin{split}
\Psi(\mbf{x})=&E(\mbf{x}-\mbf{x}_0)\Big[\psi_0\\
&+\sqrt{2}\cos(k_rz)\left(\psi_\text{F}e^{ik_rx}+\psi_\text{B}e^{-ik_rx}\right)\Big],
\end{split}\ee
where $E(\mbf{x}-\mbf{x}_0)$ is the envelope wavefunction of the BEC centered at position $\mbf{x}_0=(z_0,\mbf{r}_0)$, and $\psi_0$, $\psi_\text{F}$, and $\psi_\text{B}$ are amplitudes describing the ground-state fraction and the forward-scattered and the backward-scattered condensate, respectively. In what follows, we suppress the explicit notation of the BEC position $(\mbf{r}_0,z_0)$. The normalization condition of the wavefunction ensures that $|\psi_0|^2+|\psi_\text{F}|^2+|\psi_\text{B}|^2=1$.

We evaluate the Hamiltonian Eq.~\eqref{hamiltonian} using this ansatz. To evaluate the integrals, we assume that the extent of the envelope wavefunction is large compared to $\lambda$, allowing us to drop fast oscillating terms. We then find the atomic Hamiltonian to be
\begin{widetext}\begin{equation}\begin{split}
    H_A =&\,\, 2NE_r\left(|\psi_\text{F}|^2+|\psi_\text{B}|^2\right) + E_\text{trap}\left(|\psi_0|^2+|\psi_\text{F}|^2+|\psi_\text{B}|^2\right) \\
    &+ E_\text{int}\left[|\psi_0|^4 + 4|\psi_0|^2\left(|\psi_\text{F}|^2+|\psi_\text{B}|^2\right)+2\psi_0^2\psi_\text{F}^*\psi_\text{B}^*+2(\psi_0^*)^2\psi_\text{F}\psi_\text{B}+\frac{3}{2}\left(|\psi_\text{F}|^4+|\psi_\text{B}|^4+4|\psi_\text{F}|^2|\psi_\text{B}|^2\right)\right],
\end{split}\end{equation}\end{widetext}
where $E_r=\hbar^2k^2/(2m)\approx 3.7$~kHz is the recoil energy, $E_\text{trap} = N \int d^3\mbf{x}V(\mbf{x})E(\mbf{x})^2$ is the trap energy, and $E_\text{int} = \frac{1}{2}N^2U_0 \int d^3\mbf{x}E(\mbf{x})^4$ is the interaction energy. Using the Thomas-Fermi approximation for the envelope wavefunction, we can evaluate these integrals~\cite{Pethick2002} to be $E_\text{trap}=\frac{3}{7}\mu_\text{TF}N$ and $E_\text{int}=\frac{2}{7}\mu_\text{TF}N$, where $\mu_\text{TF}=(15\hbar^2aN\omega_x\omega_y\omega_z)^{2/5}m^{1/5}/2$ is the chemical potential for the Thomas-Fermi cloud expressed in terms of the trap frequencies $\{\omega_x,\omega_y,\omega_z\}$ and atomic properties. For the light-matter interaction, again we drop fast oscillating terms, which yields~\footnote{On the second line, we should write $\cos(\theta_\mu-\theta_\nu)=\mathcal{O}_\mu\mathcal{O}_\nu + \sin(\theta_\mu)\sin(\theta_\nu)$, but we can omit the $\sin$ terms as they all evaluate to zero for the mode indices of interest.}
\bea
H_\text{LM} &=& \frac{N\Omega^2}{\Delta_A} \\
    &&+ \frac{Ng_0\Omega}{\sqrt{2}\Delta_A}\left(\sum\limits_\mu\hat{a}_\mu I_\mu\mathcal{O}_\mu\left(\psi_0^*\psi_\text{F}+\psi_0\psi_\text{B}^*\right) + \text{H.c.}\right)\nonumber\\
    &&+\frac{Ng_0^2}{2\Delta_A}\sum\limits_{\mu,\nu}\hat{a}_\mu^\dagger\hat{a}_\nu J_{\mu,\nu}\mathcal{O}_\mu\mathcal{O}_\nu+O(\hat{a}^\dagger\hat{a}|\psi_\text{F,B}|^2),\nonumber
\eea
where $\mathcal{O}_\mu=\cos[\theta_\mu(z_0)]$, $I_\mu=\int d{r}\rho(\mbf{r})\Xi_\mu(\mbf{r})$, and $J_{\mu,\nu}=\int d{r}\rho(\mbf{r})\Xi_\mu(\mbf{r})\Xi_\nu(\mbf{r})$. $\rho(\mbf{r})=\int dz E(\mbf{x})^2$ is the transverse density distribution. We did not explicitly write out the last term containing two photon operators and two scattered states as this higher-order term will not affect the threshold condition.

To find the threshold condition, we perform a linear stability analysis of the normal state $(\psi_0,\psi_F,\psi_B)=(1,0,0)$. To do so, we need to ensure that this state is stationary, i.e., that the $\psi_0$ component has no trivial dynamics associated with it. We do this by subtracting an energy offset to the Hamiltonian of the form $\eta(|\psi_0|^2+|\psi_F|^2+|\psi_B|^2)$.  This is equivalent to adding a factor $e^{-i\eta t}$ in Eq.\eqref{eq:atomic_wf}, so all atomic components are in a frame rotating at frequency $\eta$. We then write the mean-field equations of motion given the total Hamiltonian by taking expectation values $\alpha_\mu\equiv\langle\hat{a}_\mu\rangle$:
\begin{widetext}\begin{equation}\begin{split}\label{eq:EOM}
    i\partial_t\alpha_\mu&=-(\Delta_\mu+i\kappa)\alpha_\mu + \frac{Ng_0\Omega}{\sqrt{2}\Delta_A}I_\mu\mathcal{O}_\mu\left(\psi_0\psi_F^*+\psi_0^*\psi_B\right)+\frac{Ng_0^2}{2\Delta_A}\sum\limits_\nu J_{\mu,\nu}\mathcal{O}_\mu\mathcal{O}_\nu\alpha_\nu\\
    i\partial_t\psi_0 &=\left(E_\text{trap} + E_\text{int}\left(2|\psi_0|^2 + 4|\psi_F|^2+4|\psi_B|^2\right)-\eta\right)\psi_0+4E_\text{int}\psi_0^*\psi_F\psi_B+ \frac{Ng_0\Omega}{\sqrt{2}\Delta_A}\sum\limits_\mu I_\mu\mathcal{O}_\mu\left(\alpha_\mu \psi_F+\alpha_\mu^*\psi_B\right)\\
    i\partial_t\psi_F &= \left(2NE_r+E_\text{trap}+E_\text{int}\left(4|\psi_0|^2+3|\psi_F|^2+6|\psi_B|^2\right)-\eta\right)\psi_F + 2E_\text{int}\psi_0^2\psi_B^*+ \frac{Ng_0\Omega}{\sqrt{2}\Delta_A}\psi_0\sum\limits_\mu I_\mu\mathcal{O}_\mu\alpha_\mu^*\\
    i\partial_t\psi_B &= \left(2NE_r+E_\text{trap}+E_\text{int}\left(4|\psi_0|^2+6|\psi_F|^2+3|\psi_B|^2\right)-\eta\right)\psi_B + 2E_\text{int}\psi_0^2\psi_F^*+ \frac{Ng_0\Omega}{\sqrt{2}\Delta_A}\psi_0\sum\limits_\mu I_\mu\mathcal{O}_\mu\alpha_\mu,
\end{split}\end{equation}\end{widetext}
where we introduced a nonunitary term $-i\kappa\alpha_\mu$ to account for field loss at rate $\kappa$. From inspecting the second equation, we see that picking $\eta\equiv E_\text{trap}+2E_\text{int}=\mu_{\text{TF}} N$ makes the normal state stationary~\cite{Pethick2002}. Next, since the timescale for the dynamics of the photons is much faster than the atoms, we can adiabatically eliminate the cavity fields. The last term in the first equation represents the effects of the dispersive shift, and while this term can normally be absorbed into $\Delta_\mu$, we note that in a multimode cavity it mixes different modes. Because $|Ng_0^2/(2\Delta_A)|\ll|\Delta_\mu|$, we solve the system of coupled modes by treating the coupling perturbatively through the matrix inverse approximation $(1-A)^{-1}\approx 1+A$, valid when $|\text{eig}(A)|\ll1$. Then the instantaneous steady state is
\begin{widetext}\be\label{eq:modepopulation}
\alpha_\mu=\frac{Ng_0\Omega}{\sqrt{2}\Delta_A}\left(\psi_0\psi_F^* + \psi_0^*\psi_B\right)\left[\frac{I_\mu\mathcal{O}_\mu}{\Delta_\mu+i\kappa}+\frac{Ng_0^2}{2\Delta_A}\sum\limits_\nu\frac{J_{\mu,\nu}I_\nu\mathcal{O}_\mu\mathcal{O}_\nu^2}{(\Delta_\mu+i\kappa)(\Delta_\nu+i\kappa)}\right].
\ee\end{widetext}
We insert this back into Eq.~\eqref{eq:EOM} along with the choice of $\eta$ and perform a linear stability analysis by expanding around the normal state $(\psi_0,\psi_F,\psi_B) = (1,0,0) + (\delta\psi_0,\delta\psi_F,\delta\psi_B)$ and keeping  quantities only linear in the deviations $\delta\psi$. The equation of motion for $\delta\psi_0$ is trivial, $i\partial_t\delta\psi_0 = 0$, and can be ignored. For the scattered components, we arrive at the coupled system
\begin{widetext}\begin{equation}\label{eq:stability}
    i\partial_t\begin{bmatrix} \delta\psi_F \\ \delta\psi_B \\ \delta\psi_F^* \\ \delta\psi_B^* \end{bmatrix} = \begin{bmatrix} 2NE_r+2E_\text{int} + E_\text{cav}^* & 0 & 0 & 2E_\text{int}+E_\text{cav}^* \\
    0 & 2NE_r+2E_\text{int}+E_\text{cav} & 2E_\text{int}+E_\text{cav} & 0 \\
    0 & -2E_\text{int}-E_\text{cav} & -2NE_r-2E_\text{int}-E_\text{cav} & 0 \\
    -2E_\text{int}-E_\text{cav}^* & 0 & 0 & -2NE_r - 2E_\text{int}-E_\text{cav}^*\end{bmatrix} \begin{bmatrix} \delta\psi_F \\ \delta\psi_B \\ \delta\psi_F^* \\ \delta\psi_B^* \end{bmatrix},
\end{equation}
where we introduced the shorthand
\begin{equation}\begin{split}
    E_\text{cav} &= \frac{N^2g_0^2\Omega^2}{2\Delta_A^2}\sum\limits_\mu\frac{I_\mu^2\mathcal{O}_\mu^2}{\Delta_\mu+i\kappa}+\frac{N^3g_0^4\Omega^2}{4\Delta_A^3}\sum\limits_{\mu,\nu}\frac{J_{\mu,\nu}I_\mu I_\nu\mathcal{O}_\mu^2\mathcal{O}_\nu^2}{(\Delta_\mu+i\kappa)(\Delta_\nu+i\kappa)}\\
    &= \frac{N^2g_0^2\Omega^2}{2\Delta_A^2\Delta_C}\left(\int d\mbf{r}d\mbf{r'}\rho(\mbf{r})\rho(\mbf{r'})\mathcal{D}(\mbf{r},\mbf{r'}) + \frac{Ng_0^2}{2\Delta_A}\int d\mbf{r}d\mbf{r'}d\mbf{r''}\rho(\mbf{r})\rho(\mbf{r'})\rho(\mbf{r''})\mathcal{D}(\mbf{r},\mbf{r'})\mathcal{D}(\mbf{r},\mbf{r''})\right).
\end{split}\end{equation}\end{widetext}
The last line follows from inserting the definitions of $I_\mu$ and $J_{\mu,\nu}$ and using the definition
\begin{equation}
    \mathcal{D}(\mbf{r},\mbf{r'})=\Delta_C\sum\limits_\mu\frac{\Xi_\mu(\mbf{r})\Xi_\mu(\mbf{r'})}{\Delta_\mu+i\kappa}\cos^2[\theta_\mu(z_0)].
\end{equation}
We note that this $\mathcal{D}$ is not quite the same quantity as the 3-dimensional Green's function defined in Eq.~\eqref{eq:cav_interaction}, but rather only the transverse part of it. Indeed when $z_0=0$,   $\mathcal{D}(\mbf{r},\mbf{r'})=\mathcal{D}(\mbf{r},\mbf{r'},0,0)$. 

The system given by Eq.~\eqref{eq:stability} is unstable  when any eigenvalue of the matrix accrues a (positive) imaginary part. Its eigenvalues are given by $\pm\sqrt{4NE_r(NE_r+E_\text{cav}+2E_\text{int})}$ and their complex conjugates. Since $E_\text{cav}$ is complex-valued, these eigenvalues have an imaginary part for any (nonzero) pump strength $\Omega$. However, this imaginary part is proportional to $\kappa/(\Delta_C^2+\kappa^2)$, and since we operate off-resonance where $|\Delta_C|\gg\kappa$, this corresponds to an instability with a growth rate much slower than the duration of our experiment. Instead, the superradiant threshold is observed when the real part of the argument of the square root becomes negative, resulting in the threshold condition
\begin{equation}
    -2\Re\{E_\text{cav}\} = 2NE_r + 4E_\text{int} = N\left(2E_r+\frac{8}{7}\mu\right) \equiv NE_\text{dw}.
\end{equation}
This equality is equivalent to the threshold condition presented in the main text as Eq.~\eqref{threshold_condition}.

Remembering that $\rho(\mbf{r})$ is centered at the BEC position $\mbf{r}_0$, we note that the threshold is predominantly set by the competition between the photon-mediated self-interaction energy gained at position $\mbf{r}_0$, which is approximately $\frac{N^2g_0^2\Omega^2}{\Delta_A^2\Delta_C}\mathcal{D}(\mbf{r}_0,\mbf{r}_0)$, versus the kinetic and scattering energy cost of self-organization, given by $N E_\text{dw}$. The self-interaction energy is slightly modified due to the finite extent of the cloud, hence the resulting integrals in the threshold condition. The second term in Eq.~\eqref{threshold_condition} describes the effect of the multimode dispersive shift due to the BEC at that position. 

To find the field inside the cavity that arises after the superradiant transition, we insert Eq.~\eqref{eq:modepopulation} back into the second line of Eq.~\eqref{eq:lightfield}. Ignoring the wavefront curvature again, this yields
\begin{widetext}\be\begin{split}
\Phi(\mbf{r},z)&=\frac{Ng_0^2\Omega}{\sqrt{2}\Delta_A}\left(\psi_0\psi_F^* + \psi_0^*\psi_B\right)\Bigg(\sum\limits_\mu\frac{I_\mu\mathcal{O}_\mu}{\Delta_\mu+i\kappa}\Xi_\mu(\mbf{r})\cos\left[k_rz-\theta_\mu(z)\right]\\
&\hspace{13em}+\frac{Ng_0^2}{2\Delta_A}\sum\limits_{\mu,\nu}\frac{J_{\mu,\nu}I_\nu\mathcal{O}_\mu\mathcal{O}_\nu^2}{(\Delta_\mu+i\kappa)(\Delta_\nu+i\kappa)}\Xi_\mu(\mbf{r})\cos\left[k_rz-\theta_\mu(z)\right]\Bigg)\\
&=\frac{Ng_0^2\Omega}{\sqrt{2}\Delta_A\Delta_C}\left(\psi_0\psi_F^* + \psi_0^*\psi_B\right)\bigg(\int d\mbf{r'}\rho(\mbf{r'})\mathcal{D}(\mbf{r},\mbf{r'},z,0)\\
&\hspace{14em}+\frac{Ng_0^2}{2\Delta_A\Delta_C}\int d\mbf{r'}d\mbf{r''}\rho(\mbf{r'})\rho(\mbf{r''})\mathcal{D}(\mbf{r},\mbf{r'},z,0)\mathcal{D}(\mbf{r'},\mbf{r''},0,0)\bigg),
\end{split}\ee\end{widetext}
where we again set $z_0=0$  to be able to use the 3D Green's functions. In the main text, we define the prefactor as $\Phi_0 \equiv Ng_0^2\Omega\left(\psi_0\psi_F^* + \psi_0^*\psi_B\right)/(\sqrt{2}\Delta_A\Delta_C) $, and omit the second subleading term.

\subsection{Confocal cavity Green's function}
To derive a more explicit expression for the cavity-induced interaction, we can focus on the midplane of a confocal cavity near an even mode resonance. That is, where the modes with only even $n_\mu=l_\mu+m_\mu$ participate. At this location, $\mathcal{O}_{\mu}=\cos(n_\mu\pi/4)$. As described in the main text, we model the cavity as having a linear dispersion $\Delta_\mu=\Delta_C-\epsilon n_\mu$ and an exponential mode cutoff $\exp(-\alpha n_\mu)$. The cavity interaction expression becomes
\bea
    \mathcal{D}(\mbf{r},\mbf{r'}) &=& \Delta_C\sum\limits_{\substack{\mu\\ n_\mu\in2\mathbb{N}_0}}\frac{\Xi_\mu(\mbf{r})\Xi_\mu(\mbf{r'})}{\Delta_C-\epsilon n_\mu+i\kappa}\mathcal{O}_\mu^2e^{-\alpha n_\mu} \nonumber\\ &=&\sum\limits_{\mu}\frac{\Xi_\mu(\mbf{r})\Xi_\mu(\mbf{r'})}{1+\tilde\epsilon n_\mu+i\tilde\kappa}e^{-\alpha n_\mu}\\
    &&\times\cos^2\left(\frac{n_\mu\pi}{4}\right)\cos^2\left(\frac{n_\mu\pi}{2}\right)\nonumber,
\eea
where $\tilde\epsilon=-\epsilon/\Delta_C$ and $\tilde\kappa=\kappa/\Delta_C$.  In the second line, we insert another factor of $\cos^2(n_\mu\pi/2)$ to select the appropriate even modes. To convert these sums to integrals, we can use the Green's function for Hermite-Gauss modes:
\bea
    &&G(\mbf{r},\mbf{r'},t) \equiv \sum\limits_{\mu}\Xi_\mu(\mbf{r})\Xi_\mu(\mbf{r'})t^{n_
    \mu}\\\nonumber
    &=&\frac{1}{1-t^2}\exp\left[-\frac{1+t^2}{1-t^2}\frac{\mbf{r}^2+\mbf{r'}^2}{w_0^2}+\frac{4t}{1-t^2}\frac{\mbf{r}\cdot\mbf{r'}}{w_0^2}\right].
\eea
The analytical expression on the second line can be found using the Mehler kernel for Hermite polynomials. The interaction strength can then be written as
\begin{equation}\label{eq:GreensToInteraction}
    \mathcal{D}(\mbf{r},\mbf{r'}) = \int\limits_0^\infty d{\tau}G_\text{sym}(\mbf{r},\mbf{r'},e^{-\tilde{\epsilon}\tau-\alpha})e^{-(1+i\tilde\kappa)\tau},
\end{equation}
where the mode-selecting cosines have been incorporated by symmetrizing the Green's function:
\bea\label{symmetrization}
    4G_\text{sym}(\mbf{r},\mbf{r'},t) =&& G(\mbf{r},\mbf{r'},t)+G(\mbf{r},\mbf{r'},-t)\\\nonumber &+& G(\mbf{r},\mbf{r'},it)+G(\mbf{r},\mbf{r'},-it).
\eea

As discussed above, the cavity photon-mediated atom-atom interaction is proportional to $\mathcal{D}(\mbf{r},\mbf{r'})$~\cite{Vaidya2018tpa}.  However, the interaction for two identical BECs at positions $\mbf{r}_i$ and $\mbf{r}_j$ must account for their finite size. This can be analytically performed by assuming a Gaussian for the density distribution $\rho$. While the functional form may sometimes be closer to parabolic---i.e., a Thomas-Fermi profile---we have numerically verified that using a Gaussian leads to only a very small systematic error in the measurements if the width of the Gaussian is scaled to match the Thomas-Fermi radius. The finite-size-corrected Green's function is then provided by a Gaussian integral. Direct evaluation yields
\begin{widetext}\be
G'(\mbf{r}_i,\mbf{r}_j,t)\equiv\int d\mbf{r}d\mbf{r'}\rho(\mbf{r};\mbf{r}_i)\rho(\mbf{r'};\mbf{r}_j)G(\mbf{r},\mbf{r'},t)
=\frac{(1+\gamma)^2}{4(1-\gamma^2t^2)}\exp\left(-\frac{1+\gamma}{4(1-\gamma^2t^2)}{\mbf{r}_2}\begin{bmatrix}1+\gamma t^2 & -(1+\gamma)t\\ -(1+\gamma)t & 1+\gamma t^2\end{bmatrix}{\mbf{r}_2}^T\right),
\ee\end{widetext}
where $\rho(\mbf{r};\mbf{r}_i)=\exp[-(\mbf{r}-\mbf{r}_i)^2/2 \sigma_A^2]/(2\pi\sigma_A^2)$ is the Gaussian atomic profile centered at $\mbf{r}_i$, ${\mbf{r}_2}=\sqrt{2}[\mbf{r}_i,\mbf{r}_j]/w_0$, and $\gamma=\frac{1-2\sigma_A^2/w_0^2}{1+2\sigma_A^2/w_0^2}$ captures the finite size effects of an isotropic cloud with Gaussian width $\sigma_A$. A generalized result for anisotropic gases, where $\sigma_{A,x}\neq\sigma_{A,y}$, may be similarly derived. The resulting cavity interaction can be found using the analogue of Eq.~\eqref{eq:GreensToInteraction} after symmetrizing $G'$ in the same fashion.

With regard to the last, dispersive shift term in Eq.~\eqref{threshold_condition}, we can use a similar strategy involving the Green's function and Gaussian integrals. The result for an isotropic cloud is
\begin{widetext}\bea
&&G^D(\mbf{r}_i,\mbf{r}_j,\mbf{r}_k,t,t')\equiv\int d\mbf{r}d\mbf{r'}d\mbf{r''}\rho(\mbf{r};\mbf{r}_i)\rho(\mbf{r'};\mbf{r}_j)\rho(\mbf{r''};\mbf{r}_k)G(\mbf{r},\mbf{r'},t)G(\mbf{r},\mbf{r''},t')\\\nonumber
&=&\frac{(1+\gamma)^3}{4a(t,t',\gamma)}\exp\left(-\frac{1+\gamma}{4a(t,t',\gamma)}{\mbf{r}_3}
\begin{bmatrix}
4(1-\gamma^2t^2t'^2) & -2t(1+\gamma)(1-\gamma t'^2) & -2t'(1+\gamma)(1-\gamma t^2) \\
-2t(1+\gamma)(1-\gamma t'^2) & b(t,t',\gamma) & -2(1-\gamma^2)tt' \\
-2t'(1+\gamma)(1-\gamma t^2) & -2(1-\gamma^2)tt' & b(t',t,\gamma)
\end{bmatrix}{\mbf{r}_3}^T\right),
\eea
where $\mbf{r}_3=\sqrt{2}[\mbf{r}_i,\mbf{r}_j,\mbf{r}_k]/w_0$ and 
\bea
a(t,t',\gamma)&=&3 - \gamma(1+t^2+t'^2)-\gamma^2(t^2+t'^2+t^2t'^2)+3\gamma^3t^2t'^2\\
b(t,t',\gamma)&=&3+t^2-\gamma(1-t^2)(1+t'^2)-\gamma^2t'^2(1+3t^2).
\eea\end{widetext}
Again this result can be generalized to the case of anisotropic clouds.

Finally, the threshold condition can be explicitly written in terms of the Green's functions:
\begin{widetext}\bea\label{eq:threshold_full}
E_\text{dw}&=&-\frac{Ng_0^2\Omega_\text{c}^2}{\Delta_A^2\Delta_C}\Re\Bigg\{\int\limits_0^\infty d\tau G'_\text{sym}(\mbf{r}_0,\mbf{r}_0,e^{-\tilde\epsilon\tau-\alpha})e^{-(1+i\tilde\kappa)\tau}\\\nonumber &&\hspace{7em}+\frac{Ng_0^2}{2\Delta_A\Delta_C}\int\limits_0^\infty d\tau \int\limits_0^\infty d\lambda G^D_\text{sym}(\mbf{r}_0,\mbf{r}_0,\mbf{r}_0,e^{-\tilde\epsilon\tau-\alpha},e^{-\tilde\epsilon\lambda-\alpha})e^{-(1+i\tilde\kappa)\tau}e^{-(1+i\tilde\kappa)\lambda}\Bigg\}\\\nonumber
&=&-\frac{Ng_0^2\Omega_\text{c}^2}{\Delta_A^2\Delta_C}\frac{e^{u\alpha}}{\tilde\epsilon}\Re\left\{\int\limits_0^{e^{-\alpha}}dt G'_\text{sym}(\mbf{r}_0,\mbf{r}_0,t)t^{u-1}+\frac{Ng_0^2}{2\Delta_A\Delta_C}\frac{e^{u\alpha}}{\tilde\epsilon}\int\limits_0^{e^{-\alpha}}dt\int\limits_0^{e^{-\alpha}}dt' G^D_\text{sym}(\mbf{r}_0,\mbf{r}_0,\mbf{r}_0,t,t')(tt')^{u-1}\right\},
\eea\end{widetext}
where $G^D_\text{sym}$ is symmetrized according to Eq.~\eqref{symmetrization} with respect to both $t$ and $t'$ arguments and $u\equiv (1+i\tilde{\kappa})/\tilde{\epsilon}$. The last line, which follows from a change of the integration variables to $t$ and $t'$, offers a more convenient expression because it is amenable to numerical evaluation using quadrature techniques~\cite{Davis2007mon}. We use a midpoint rule on an inhomogeneous grid to evaluate the threshold for the fits.

\subsection{Cooperativity enhancement}\label{Sec:Enhancement}
The threshold condition for a single-mode cavity is given by
\begin{equation}
    E_\text{dw}=-\frac{Ng_0^2\Omega_c^2}{\Delta_A^2\Delta_C}\Re\left\{\int d\mbf{r}d\mbf{r'}\rho(\mbf{r})\rho(\mbf{r'})\frac{\Xi_{00}(\mbf{r})\Xi_{00}(\mbf{r'})}{1+i\tilde\kappa}\right\}.
\end{equation}
Comparing to the multimode threshold condition Eq.~\eqref{threshold_condition}, we note that the (near-)degenerate presence of higher order modes lowers the threshold by a factor of
\begin{equation}
\frac{C_\text{mm}}{C}=\frac{\Re\left\{\int d\mbf{r}d\mbf{r'}\rho(\mbf{r})\rho(\mbf{r'})\mathcal{D}(\mbf{r},\mbf{r'})\right\}}{\Re\left\{\int d\mbf{r}d\mbf{r'}\rho(\mbf{r})\rho(\mbf{r'})\frac{\Xi_{00}(\mbf{r})\Xi_{00}(\mbf{r'})}{1+i\tilde\kappa}\right\}}.
\end{equation}

To find the cavity-limited enhancement, we first take the limit of a point particle~\footnote{$N=1$ for a point particle, hence we omitted the dispersive shift term from Eq.~\eqref{threshold_condition} in the above expression for $C_\text{mm}$ because its contribution is negligible.} placed at the cavity center where the light-matter coupling is strongest, $\rho(\mbf{r})=\delta(\mbf{r})$. We then find  that $C_\text{mm}/C=\Re\{\mathcal{D}(\mbf{0},\mbf{0})\}/\Re\{\frac{1}{1+i\tilde\kappa}\}\approx\mathcal{D}(\mbf{0},\mbf{0})$, because $\tilde\kappa\ll1$. Using the weights $W_\mu = (1+\tilde\epsilon n_\mu + i\tilde\kappa)^{-1}e^{-\alpha n_\mu}$, this can be expressed as 
\begin{equation}
\frac{C_\text{mm}}{C}=\frac{1}{4\tilde\epsilon}\mathcal{P}\left(e^{-4\alpha},1,\frac{u}{4}\right),
\end{equation}
where $\mathcal{P}(z,s,a)=\sum\limits_{n=0}^\infty z^n(n+a)^{-s}$ is the Lerch transcendent~\cite{Apostol2010zar}. 

An extended gas at the cavity center experiences a reduced cooperativity enhancement given by $\int d\mbf{r}d\mbf{r'}\rho(\mbf{r})\rho(\mbf{r'})\mathcal{D}(\mbf{r},\mbf{r'})$. Using $G'_\text{sym}$, we find a similar expression to before,
\begin{equation}
\frac{C_\text{mm}^\text{cloud}}{C}=\frac{(1+\gamma)^2}{16\tilde\epsilon}\mathcal{P}\left(\gamma^4e^{-4\alpha},1,\frac{u}{4}\right).
\end{equation}
It is worth noting that other than occurring as the prefactor, $\gamma$ renormalizes $\alpha\rightarrow\alpha-\ln(\gamma)$. This makes accurate determination of $\alpha$ harder, as finite size obfuscates it. Finally, if the gas is anisotropic, described by $\gamma_{x(y)} = \frac{1-2\sigma_{A,x(y)}^2/w_0^2}{1+2\sigma_{A,x(y)}^2/w_0^2}$ for the $x$ ($y$) direction, we can find the cooperativity enhancement as
\begin{widetext}\begin{equation}
\frac{C_\text{mm}^\text{aniso}}{C} = \frac{(1+\gamma_x)(1+\gamma_y)}{8u\tilde\epsilon}\left[F_1\left(\frac{u}{2},\frac{1}{2},\frac{1}{2},1+\frac{u}{2},-\gamma_x^2e^{-2\alpha},-\gamma_y^2e^{-2\alpha}\right)+F_1\left(\frac{u}{2},\frac{1}{2},\frac{1}{2},1+\frac{u}{2},\gamma_x^2e^{-2\alpha},\gamma_y^2e^{-2\alpha}\right)\right],
\end{equation}\end{widetext}
where $F_1$ is the Appell hypergeometric function~\cite{Askey2010ghf}. This is the dominant contribution to the theory curves shown in Fig.~\ref{fig3}a, with the remainder coming from the dispersive shift term.

Note that we did not define the multimode cooperativity through a first-principles approach that relates it to the mode volume of $\Phi(\mbf{r},z)$. The main reason is that in a multimode cavity with imperfect degeneracy, $\Phi(\mbf{r},z)$ represents the field of the \textit{synthetic} mode:  It does not exist as an eigenmode at a single optical frequency, due to the dephasing from the dispersion of the constituent bare cavity modes.  That is, this synthetic mode only exists as the photonic part of a coupled light-matter state when the pumping is present, as discussed earlier.  Hence, a mode volume calculation yields an incorrect answer for the enhancement. This subtlety between synthetic mode and supermode disappears when the modes are exactly degenerate. Indeed, a mode volume calculation agrees with the above in that case.

\subsection{Atom interaction as a convolution}\label{convolution}
The position scans as presented in the main text can be approximately understood as a convolution between the local part of the cavity-mediated interaction and a density kernel defined as $\rho_2(\mbf{r})=\int d\mbf{r'}\rho(\mbf{r'})\rho(\mbf{r}-\mbf{r'})$~\cite{Vaidya2018tpa}. Ignoring the nonlocal interaction, we can write $\mathcal{D}(\mbf{r},\mbf{r'})\approx g(\mbf{r}-\mbf{r'}) + g(\mbf{r}+\mbf{r'})$, where $g(\Delta\mbf{r})$ is the functional form of the local interaction and is (approximately) translationally invariant;  the term $g(\mbf{r}+\mbf{r'})$ captures the ``mirror'' image that arises because of the mirror symmetry at an even resonance. Plugging this in and assuming that $\rho(\mbf{r})$ is symmetric around $\mbf{r}=\mbf{r}_0$, we obtain  the first integral in the threshold expression Eq.~\eqref{eq:threshold_full}:
\begin{equation}\begin{split}
    \int d\mbf{r}d\mbf{r'}\rho(\mbf{r})&\rho(\mbf{r'})\mathcal{D}(\mbf{r},\mbf{r'})\\&\approx\int d\mbf{z}\rho_2(\mbf{z})g(\mbf{z}) + \int d\mbf{z}\rho_2(\mbf{z})g(2\mbf{r}_0-\mbf{z})\\
    &=h(\mbf{0}) + h(2\mbf{r}_0),
\end{split}\end{equation}
where $h(\mbf{r})=\int d\mbf{z}\rho_2(\mbf{z})g(\mbf{r}-\mbf{z})$ is also a convolution.

\section{Fitting procedure}\label{fitprocedure}
To fit the superradiant threshold data, we rewrite the threshold expression Eq.~\eqref{threshold_condition} as
\begin{equation}
    \frac{E_\text{dw}}{N\Omega_\text{c}^2} = -\frac{g_0^2}{\Delta_A^2\Delta_C}\int d\mbf{r}\rho(\mbf{r})\Re\left\{\frac{\Phi(\mbf{r})}{\Phi_0}+\frac{Ng_0^2}{2\Delta_A\Delta_C}\frac{\Phi(\mbf{r})^2}{\Phi_0^2}\right\}.
\end{equation}
The left-hand side of this equation consists of the observed threshold pump power $\Omega_c^2$ and atom number at threshold $N$, and forms the dependent variable $y$. The right-hand side depends on the independent variables $x=\{\mbf{r}_0,R_x,R_y,\Delta_C\}$, the constants $c=\{g_0,\Delta_A\}$, and the fitting parameters $p=\{\epsilon,\alpha,\Delta_0,\{A_i\}\}$. $\epsilon$ and $\alpha$ come in through the weighting factors $W_\mu$ of the cavity Green's function as explained in the main text. $\Delta_0$ is an additional offset applied to all cavity detunings because in the nonideal confocal cavity, resonance is not clearly defined. All $\Delta_C$ are recorded with respect to a fixed reference point in the cavity transmission spectrum as described in the main text, and $\Delta_0$ accounts for mismatch between the arbitrarily chosen reference point and the effective zero detuning point. The $\{A_i\}$ are overall multiplicative factors for each dataset; given the 18 scans and 8 on-center datasets there are a total of 26 such factors. They account for global systematics in the values of $\Omega$ and $g_0$, as well as systematics between the various datasets in determining atom number and position. With regards to positioning, each individual scan is corrected for small amounts of residual drift in the BEC position relative to cavity center by fitting the peak with an offset Gaussian and subtracting the offset.  (Figure~\ref{fig2}b is made of four such individual scans.)  These corrections are typically 1~$\mu$m or less. For these datasets, $A$ still accounts for offsets in the orthogonal direction of similar magnitude. 

For the data analysis, we perform a simultaneous least squares fit to the entire data set to find the optimal fit parameters $p^*$ and their covariance. That is, the fit simultaneously includes all the data taken with various BEC positions and sizes, as well as both of the modalities of measurement presented in Figs.~\ref{fig2}a and~\ref{fig2}b, respectively. The optimal parameters are $\{\epsilon/2\pi,\alpha,\Delta_0/2\pi\}=\{2.6(1.6)$~MHz$,0(2)\times10^{-4},0.8(18.0)$~MHz$\}$ respectively, and the typical amplitude factor is 1.4(4). Note that negative $\alpha$ is unphysical, and its reported uncertainty is a one-sided standard deviation.

To crosscheck our error estimates, we perform a bootstrap error analysis to ascertain the accuracy of the aforementioned fit.  To do so, we draw from the data (with replacement) a random sample of the same total number of data points.  The optimal fit parameters are then found again, repeating the process 300 times.  We calculate the mean and covariances from the collection of sets of fit parameters. The results are $\epsilon/2\pi=2.7(1.6)$~MHz and $\alpha=1.4(3.2)\times10^{-3}$, where the latter is heavily skewed by a small number of outliers (the median $\alpha$ is zero). As we discuss in the main text, our assessment of $\alpha$ is limited by the finite size of the BEC used. In conclusion, the bootstrap estimates of $\epsilon$ and the remaining fit parameters are very similar to the fit results reported in the main text.

\section{All-optical measurement}\label{alloptical}

\subsection{Longitudinal pumping theory}
In this section, we show that the steady-state cavity response to a longitudinal drive involves the same Green's function that describes the cavity-mediated atom-atom interaction. The Hamiltonian that describes the longitudinal pumping of a multimode cavity is 
\begin{equation}
 H = -\sum\limits_\mu\Delta_\mu\hat{a}_\mu^\dagger\hat{a}_\mu + i\kappa(f_\mu\hat{a}_\mu^\dagger - f_\mu^* \hat{a}_\mu),
\end{equation}
where
\begin{equation}
f_\mu = \int d\textbf{r}' \Xi_\mu(\textbf{r}')E_p(\textbf{r}')
\end{equation}
describes the overlap between the longitudinal pump field $E_p$ and the cavity field $\Xi_\mu(\textbf{r})$ at the midplane. Taking into account the cavity photon loss, the expectation values $\alpha_\mu = \langle \hat{a}_\mu\rangle$ satisfy the following equations of motion:
\begin{equation}
    i\partial_t \alpha_\mu = -(\Delta_\mu + i\kappa)\alpha_\mu + i\kappa f_\mu.
\end{equation}
The steady-state photon fields are derived by setting $\partial_t \alpha_\mu = 0$, which yields
\begin{equation}
    \alpha_\mu = \frac{ i\kappa f_\mu}{\Delta_\mu+i\kappa}. 
\end{equation}
The transverse cavity field can be obtained by summing over the amplitudes of the eigenmodes:
\begin{equation}
\begin{split}
    \Phi(\textbf{r}) &= \sum_\mu \alpha_\mu \Phi_\mu(\textbf{r},z = 0)\\ 
    &= i\kappa\int d\textbf{r}'\sum_\mu \frac{\Xi_\mu(\textbf{r})\Xi_\mu(\textbf{r}')}{\Delta_\mu+i\kappa}E_p(\textbf{r}') \\
    &= i\kappa\int d\textbf{r}'\mathcal{D}(\textbf{r},\textbf{r}') E_p(\textbf{r}').
\end{split}
\end{equation}
This is the same Green's function that describes cavity-mediated interactions between atoms---it has the same form as Eq.~\eqref{eq:cavfield}.  Hence, following the same logic as in Sec.~\ref{convolution}, $\Phi$ is a convolution of the cavity Green's function $\mathcal{D}$ and the longitudinal pump field $E_p$.

\subsection{Imaging effects due to cavity substrates}\label{abberations}

The light field recorded on the camera provides us with the information about the cavity field $\Phi$. However, care must be taken to distill the effect of the Green's function from other imaging aberrations arising from downstream optics. The refraction of the cavity field traveling through a cavity mirror is significant due to its thickness and small radius of curvature. In this work, we are primarily concerned with the on-axis Green's function and so our measurements concentrate on optical spots placed on the mirror substrate near the symmetric axis of the cavity.  The major mirror-induced effects are: 1) paraxial lensing that results in demagnification of the field; and 2) aberration effects of the post-cavity imaging optics, which modify the point spread function to broaden the final image. The field measured at the camera plane can be written as:
\begin{equation}
E_\text{ccd} = \text{PSF}_\text{pc}*(M[\Phi]) = \text{PSF}_\text{pc}*(M[\mathcal{D}*E_p]), 
\label{eq:pumpfieldtoccd}
\end{equation}
where $*$ denotes convolution and $M[x]$ denotes paraxial lensing that transforms $x \rightarrow mx$. $m\approx 0.69$ is the magnification calculated using a paraxial treatment of the cavity substrate. The point spread function $\text{PSF}_\text{pc}$ incorporates all the aberration effects from the post-cavity imaging optics.

In this analysis, we measure only the intensity of the cavity field, $|E_\text{ccd}|^2$, rather than the phase. More could be gleaned by detecting the electric field $E_\text{ccd}$ using holographic techniques~\cite{Schine2019eag,Kroeze2018sso,Guo2019spa}, which we leave for future work. To simplify the analysis of the detected intensity, we approximate $\text{PSF}_\text{pc}$, $\mathcal{D}$, and $E_p$ as Gaussians, so that the width of the Green's function can be estimated by taking a quadrature difference:
\begin{equation}
    \sigma(\mathcal{D})^2 = \frac{[\sigma(E_{ccd})^2-\sigma(\text{PSF}_\text{pc})^2]}{m^2}-\sigma(E_p)^2. 
\label{eq:greenquadraturesum}
\end{equation}
For $\text{PSF}_\text{pc}$, we  take into account only on-axis spherical aberrations.  This places a lower bound on $\sigma(\text{PSF}_\text{c})$. Hence, the results presented in the main text place a conservative upper bound on $\sigma(\mathcal{D})$ and the cavity resolution.

\section{Effective mode number}\label{effectivemodenumber}
We now quantify the effective number of cavity eigenmodes participating in the synthetic mode.  To do so, we can use a model with exact degeneracy but a hard cutoff, $l,m\leq M$, where we assume $M$ is even for convenience. A total of $\frac{1}{4}(M+1)^2$ modes can then participate in the synthetic mode. To find the effective value of $M$, we match the cooperativity enhancement of this model to that observed. The former can be calculated by explicitly summing the Hermite-Gauss modes at the cavity center while accounting for the mode selection via 
\begin{equation}
W_{l,m} = \begin{cases}1 &\mbox{if $l+m = 0\mod 4$,}\\
0&\mbox{otherwise.}\end{cases}
\end{equation}
In accordance with Eq.~\eqref{eq:cenhancement} this gives
\be\begin{split}
\frac{C_\text{square}}{C} &= \sum\limits_{l=0}^M\sum\limits_{m=0}^M \Xi_{l,m}(\mbf{0})^2W_{l,m}\\
&=2^{-2M-1}\frac{(M+1)!^2}{(M/2)!^4}\\
&\hspace{1em}+2^{-M-1}\frac{(M+1)!!^2}{(M/2)!^2}{}_2F_1\left(-\frac{M}{2},\frac{1}{2},\frac{3}{2},2\right)^2\\
&\approx\frac{1}{\pi}M+\frac{2}{\pi}+\sqrt{2}+\mathcal{O}(M^{-1}),
\end{split}\ee
where ${}_2F_1$ is a hypergeometric function~\cite{Askey2010ghf}, and the approximation holds for large $M$. Thus, to match the highest observed cooperativity enhancement ratio of $C_\text{mm}/C\approx 21 $, we need $M\approx\pi C_\text{mm}/C\approx 66 $, and the number of participating modes is $\frac{1}{4}(M+1)^2\approx1100$. We note that while this effective number of modes depends on the exact form of the cutoff used, the order of magnitude and scaling with $C_\text{mm}$ are generic.


%

\renewcommand{\thefigure}{S\arabic{figure}}
\setcounter{figure}{0}

\onecolumngrid
\section*{Supplemental Material: \\ A high-cooperativity confocal cavity QED microscope}

We provide all data used in the global fit here, as we presented only a small subset of all the data in the main text for clarity. As discussed in the main text, we employ BECs with various shapes and densities to suppress systematic effects. 

\begin{figure}[h!]
    \centering
    \includegraphics[width=0.99\columnwidth]{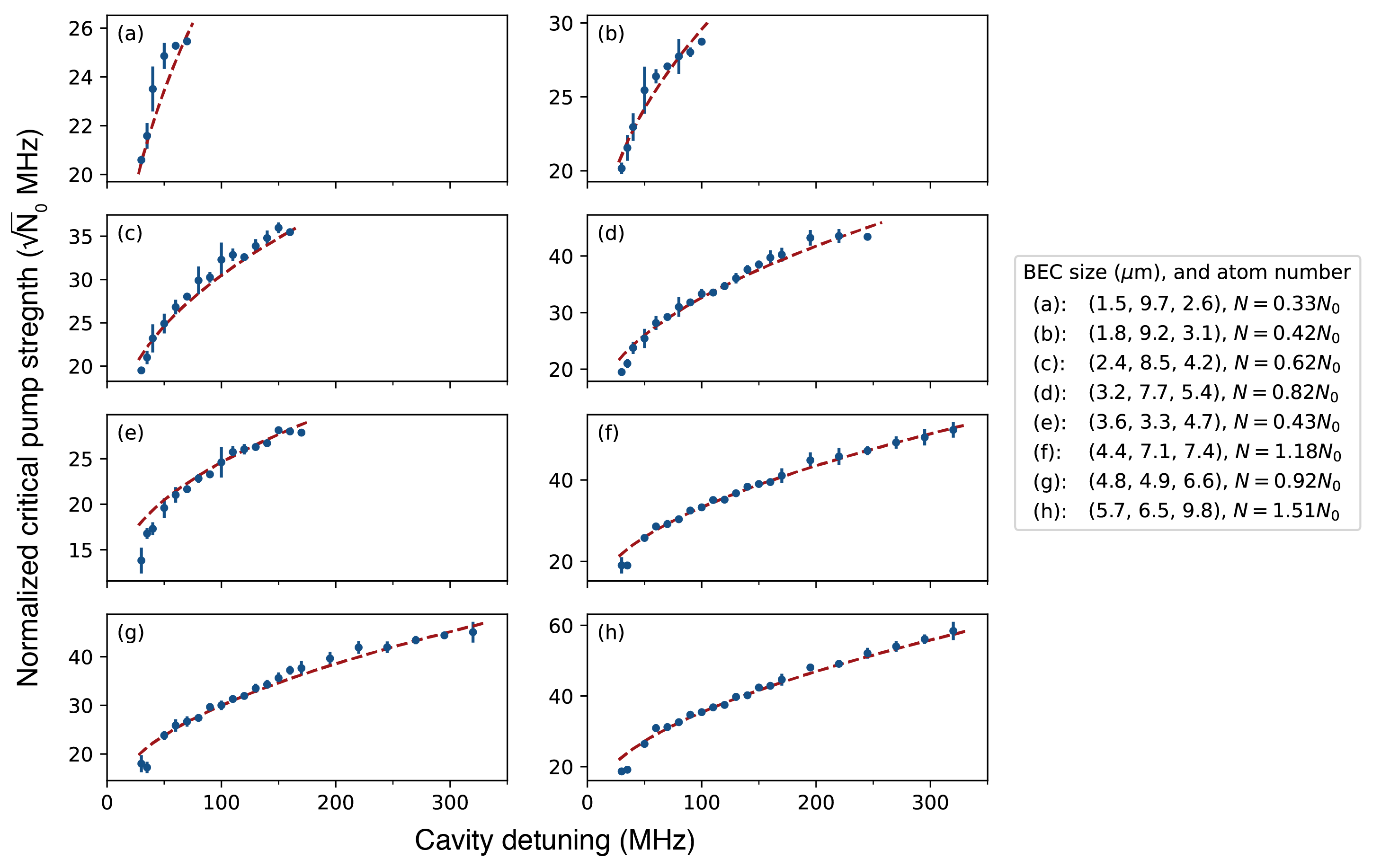}
    \caption{All of the threshold measurements taken at cavity center. The various BEC sizes are noted in the legend. Error bars indicate standard deviation over repeated measurements---typically 5 repetitions. Red dashed lines are from the global fit. The data shown in Fig.~\ref{fig2}a of the main text correspond to panels (d), (e), and (g). \label{suppfig1}}
\end{figure}

In addition to the three BEC shapes shown in Fig.~\ref{fig2}a, we used five other  BECs. These are all shown in Fig.~\ref{suppfig1}. The red dashed lines are from the global fit described in the main text.  These show good agreement with the data across all datasets.

\begin{figure}[t!]
    \centering
    \includegraphics[width=0.99\columnwidth]{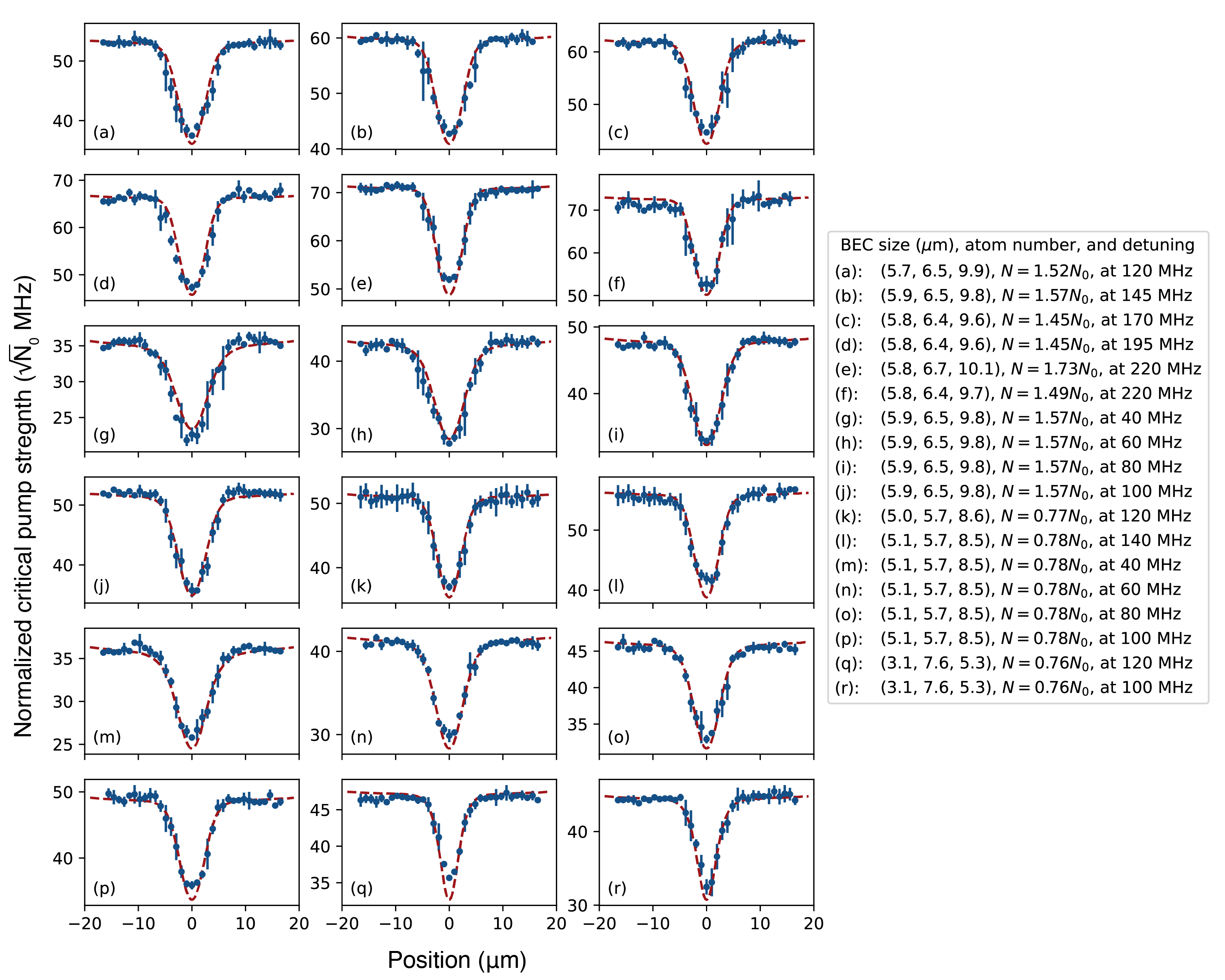}
    \caption{All of the position-dependent threshold measurements.  The various BEC sizes, atom numbers, and cavity detunings are noted in the legend. Error bars indicate standard deviation over repeated measurements---typically 4 repetitions. Red dashed lines are from the global fit. Figure~\ref{fig2}b of the main text corresponds to (q). \label{suppfig2}}
\end{figure}

In Fig.~\ref{fig2}b, we observe the threshold pump power as function of position inside the cavity. This kind of measurement was repeated for 18 different BEC shapes and cavity detunings. All these are shown in Fig.~\ref{suppfig2}. Again, the red dashed lines from the global fit are in agreement with the data.

\end{document}